\documentclass{jfm}

\usepackage{graphicx}
\usepackage{natbib}
\usepackage{amsmath}
\ifCUPmtlplainloaded \else
  \checkfont{eurm10}
  \iffontfound
    \IfFileExists{upmath.sty}
      {\typeout{^^JFound AMS Euler Roman fonts on the system,
                   using the 'upmath' package.^^J}%
       \usepackage{upmath}}
      {\typeout{^^JFound AMS Euler Roman fonts on the system, but you
                   dont seem to have the}%
       \typeout{'upmath' package installed. JFM.cls can take advantage
                 of these fonts,^^Jif you use 'upmath' package.^^J}%
       \providecommand\upi{\pi}%
      }
  \else
    \providecommand\upi{\pi}%
  \fi
\fi

\ifCUPmtlplainloaded \else
  \checkfont{msam10}
  \iffontfound
    \IfFileExists{amssymb.sty}
      {\typeout{^^JFound AMS Symbol fonts on the system, using the
                'amssymb' package.^^J}%
       \usepackage{amssymb}%
       \let\le=\leqslant  \let\leq=\leqslant
       \let\ge=\geqslant  \let\geq=\geqslant
      }{}
  \fi
\fi

\ifCUPmtlplainloaded \else
  \IfFileExists{amsbsy.sty}
    {\typeout{^^JFound the 'amsbsy' package on the system, using it.^^J}%
     \usepackage{amsbsy}}
    {\providecommand\boldsymbol[1]{\mbox{\boldmath $##1$}}}
\fi

\providecommand\bnabla{\boldsymbol{\nabla}}
\providecommand\bcdot{\boldsymbol{\cdot}}

\newcommand\etb{\boldsymbol{\eta}}

\newcommand\Real{\mbox{Re}} 
\newcommand\Imag{\mbox{Im}} 
\newcommand\Rey{\mbox{\textit{Re}}}  
\newcommand\Pran{\mbox{\textit{Pr}}} 
\newcommand\Pen{\mbox{\textit{Pe}}}  
\newcommand\Ai{\mbox{Ai}}            
\newcommand\Bi{\mbox{Bi}}            

\newcommand\slsQ{\mathsfbi{Q}} 

\newcommand\hatR{\skew3\hat{R}}      

\newsavebox{\astrutbox}
\sbox{\astrutbox}{\rule[-5pt]{0pt}{20pt}}
\newcommand{\astrut}{\usebox{\astrutbox}}

\newcommand\p{\ensuremath{\partial}}

\newcommand\kgd{\ensuremath{k\gamma d}}
\newcommand\shalf{\ensuremath{{\scriptstyle\frac{1}{2}}}}

\newcommand\squart{\ensuremath{{\textstyle\frac{1}{4}}}}
\newcommand\thalf{\ensuremath{{\textstyle\frac{1}{2}}}}
\newcommand\Gat{\ensuremath{\widetilde{G_a}}}
\newcommand\ttz{\ensuremath{\rightarrow 0}}
\newcommand\ndq{\ensuremath{\frac{\mbox{$\partial$}}{\mbox{$\partial$} n_q}}}
\newcommand\sumjm{\ensuremath{\sum_{j=1}^{M}}}
\newcommand\pvi{\ensuremath{\int_0^{\infty}%
  \mskip \ifCUPmtlplainloaded -30mu\else -33mu\fi -\quad}}

\newcommand\etal{\mbox{\textit{et al.}}}

\usepackage{subfig}
\usepackage[abs]{overpic}
\usepackage{placeins}

\usepackage{tikz}
\newcommand*\mycirc[1]{%
  \begin{tikzpicture}[baseline=(C.base)]
    \node[draw,circle,inner sep=0.2pt](C) {#1};
  \end{tikzpicture}}

\title[Non-local clustering mechanism and preferential concentration]{On the relationship between the non-local clustering mechanism and preferential concentration}

\author[A. D. Bragg, P. J. Ireland and L. R. Collins]%
{A\ls N\ls D\ls R\ls E\ls W\ns D.\ns B\ls R\ls A\ls G\ls G$^1$%
  \thanks{Email address for correspondence: adb265@cornell.edu},\ns
P\ls E\ls T\ls E\ls R\ns J.\ns I\ls R\ls E\ls L\ls A\ls N\ls D$^1$\\
\ns\and\ns L\ls A\ls N\ls C\ls E\ns R.\ns C\ls O\ls L\ls L\ls I\ls N\ls S$^1$}

\affiliation{$^1$ Sibley School of Mechanical \& Aerospace Engineering, Cornell University, Ithaca, NY 14853\\[\affilskip]}

\pubyear{2010}
\volume{650}
\pagerange{119--126}
\date{?; revised ?; accepted ?. - To be entered by editorial office}
\begin{document}

\maketitle

\begin{abstract}
`Preferential concentration' (\emph{Phys. Fluids} \textbf{A3}:1169--78, 1991) refers to the clustering of inertial particles in the high-strain, low-rotation regions of turbulence. The `centrifuge mechanism' of Maxey (\emph{J. Fluid Mech.} \textbf{174}:441--65, 1987) appears to explain this phenomenon. In a recent paper, Bragg \& Collins (\emph{New J. Phys.} \textbf{16}:055013, 2014) showed that the centrifuge mechanism is dominant only in the regime ${St\ll1}$, where $St$ is the Stokes number based on the Kolmogorov time scale. Outside this regime, the centrifuge mechanism gives way to a non-local, path-history symmetry breaking mechanism. However, despite the change in the clustering mechanism, the instantaneous particle positions continue to correlate with high-strain, low-rotation regions of the turbulence.  In this paper, we analyze the exact equation governing the radial distribution function and show how the non-local clustering mechanism is influenced by, but not dependent upon, the preferential sampling of the fluid velocity gradient tensor along the particle path-histories in such a way as to generate a bias for clustering in high-strain regions of the turbulence.  We also show how the non-local mechanism still generates clustering, but without preferential concentration, in the limit where the timescales of the fluid velocity gradient tensor measured along the inertial particle trajectories approaches zero (such as white-noise flows or for particles in turbulence settling under strong gravity).  Finally, we use data from a direct numerical simulation of inertial particles suspended in Navier-Stokes turbulence to validate the arguments presented in this study.

\end{abstract}


\section{Introduction}

Inertial particles that are initially uniformly distributed throughout an incompressible turbulent flow will develop dynamically evolving spatial clusters.  Early studies observed that inertial particles not only cluster but preferentially concentrate in strain-dominated regions of the turbulent fluid velocity field, away from regions of strong rotation \citep{maxey86,maxey87,squires91a,eaton94,wang93,sundaram4}.  Here and throughout we make a distinction between the terms `clustering' and `preferential concentration,' where the former refers to the non-uniform distribution of the particles in space, irrespective of where they cluster, and the latter refers to the clustering of particles in high-strain, low-rotation regions of the flow. The reason for making this distinction is that clustering can occur without preferential concentration, a scenario that we discuss later in the paper.

The traditional explanation for the mechanism of clustering is due to \cite{maxey87}, who argued that particles are centrifuged out of regions where the fluid streamlines exhibit strong curvature into regions of high-strain. This physical argument, aside from being intuitive, explains the preferential concentration effect. However, over time the completeness of this explanation for clustering has been questioned.  For example, \cite{bec03} described simulations of particles in white-in-time random flows that exhibited strong clustering, yet the centrifuge mechanism does not operate in such flows.

In a recent paper, \cite{bragg14b} considered in detail the physical mechanism responsible for the clustering of inertial particles in the dissipation range of isotropic turbulence.  We showed that in the regime $St\ll1$ (where $St\equiv\tau_p/\tau_\eta$ is the Stokes number, $\tau_p$ is the particle response time and $\tau_\eta$ is the Kolmogorov timescale), the mechanism for clustering predicted by the Zaichik \& Alipchenkov theory \citep[][hereafter ZT]{zaichik03,zaichik07,zaichik09} is the same as that predicted by the Chun \emph{et al.} theory \citep[][hereafter CT]{chun05}, which is essentially an extension of the classical centrifuge mechanism of \cite{maxey87}.  As $St$ is increased beyond this regime (i.e., ${St\geq O(1)}$), the system undergoes a bifurcation and the particle velocity dynamics become non-local (by `non-local' we mean that the particle velocity at a given time depends upon its interactions with the turbulent fluid velocity field at earlier times along its path-history), and we showed that this non-locality contributes to the inward drift mechanism described by the ZT through a path-history symmetry breaking effect. Figure~\ref{Non_local_mechanism_diag} illustrates two particle pairs passing through separation $\boldsymbol{r}$ at time $t$. For ${St\geq O(1)}$, the particles retain a finite memory of their interaction with the turbulence along their path histories, over a timeframe ${O(t-\tau_p)\leq s\leq t}$. The particle pair separation is given by $\boldsymbol{r}^p(s\vert\boldsymbol{r},t)$, where $\boldsymbol{r}^p$ is the time dependent separation, and $\vert\boldsymbol{r},t$ denotes that the pair has separation $\boldsymbol{r}$ at time $t$.  For the solid-line trajectory in figure~\ref{Non_local_mechanism_diag}, $\boldsymbol{r}^p(s\vert\boldsymbol{r},t)\geq\boldsymbol{r}$ and for the dashed-line trajectory, $\boldsymbol{r}^p(s\vert\boldsymbol{r},t)\leq\boldsymbol{r}$.  Since the fluid velocity difference, $\Delta\boldsymbol{u}$, on average increases with separation then particle pairs with $\boldsymbol{r}^p(s\vert\boldsymbol{r},t)\geq\boldsymbol{r}$ will have experienced, on average, larger values of $\Delta\boldsymbol{u}$ in their path-history as compared with particle pairs with $\boldsymbol{r}^p(s\vert\boldsymbol{r},t)\leq\boldsymbol{r}$.  This asymmetry in the particle pairs' path histories breaks the symmetry of their inward and outward motions and gives rise to a net inward drift, leading to clustering. 

We also argued that the theory for clustering in \cite{gustavsson11b} is qualitatively similar to that in the ZT in the regimes ${St\ll1}$ and ${St\gg1}$. The non-local, path-history symmetry breaking mechanism in the ZT and the non-local `ergodic mechanism' in \cite{gustavsson11b} both explain the clustering in white-in-time flows observed in \cite{bec03}.  We will return to this point later in the paper. Note that the ZT is more general in its description, as it applies to all $St$ and correlation timescales for the fluid, whereas \cite{gustavsson11b} treat these timescales as a small parameter.

There is however an outstanding issue that remains to be explained. \cite{bragg14b} showed that the non-local clustering mechanism in the ZT makes the dominant contribution when ${{St=O(1)}}$ \citep[see figure~3 in][and the surrounding discussion]{bragg14b}.  However, we also know from numerical simulations \citep[e.g.,][]{squires91a} that at ${{St=O(1)}}$ there exist strong correlations between the location of the clusters and the local topology of the fluid velocity gradient field, i.e., preferential concentration. In the regime ${St\ll1}$, the phenomenon of preferential concentration is easily explained since the mechanism generating the clustering depends upon the preferential sampling of the local fluid velocity gradient \citep{chun05,balachandar10}. However, for ${{St=O(1)}}$ the clustering mechanism is strongly non-local, and this symmetry-breaking clustering mechanism does not explicitly depend upon the local flow topology. How then does the non-local clustering mechanism continue to generate preferential concentration?
{\begin{figure}
\center
{\begin{overpic}
[trim = 0mm 100mm 20mm 20mm,scale=0.5,clip,tics=20]{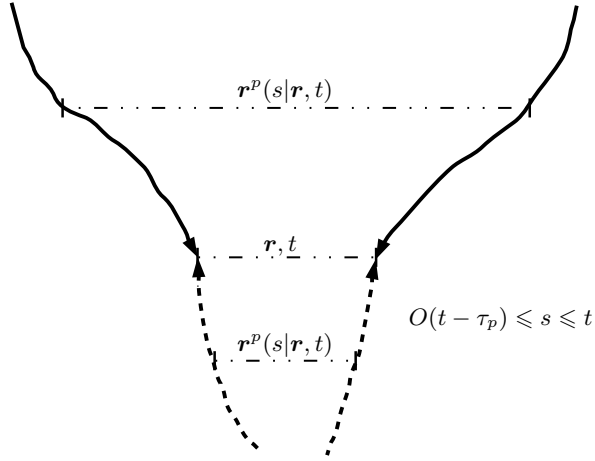}
\put(125,158){$\boldsymbol{r}^p(s\vert\boldsymbol{r},t)$}
\put(135,100){$\boldsymbol{r},t$}
\put(125,62){$\boldsymbol{r}^p(s\vert\boldsymbol{r},t)$}
\put(190,72){$O(t-\tau_p)\leq s\leq t$}
\end{overpic}}
\caption{Illustration of two particle paths arriving at separation $\boldsymbol{r}$ at time $t$. The solid lines represent a pair of particle trajectories with separation $\boldsymbol{r}^p(s\vert\boldsymbol{r},t)\geq\boldsymbol{r}$ and the dashed lines a pair with $\boldsymbol{r}^p(s\vert\boldsymbol{r},t)\leq\boldsymbol{r}$.}
\label{Non_local_mechanism_diag} 
\end{figure}
In this paper, we address this question through an analysis of the exact equation for the radial distribution function (RDF), an important measure of clustering. In particular, we resolve the issue by showing that the non-local clustering mechanism is influenced by the preferential sampling of $\boldsymbol{\Gamma}$, the fluid velocity gradient tensor, along the path histories of the particles in such a way as to generate stronger inward drift velocities into regions where $\boldsymbol{\Gamma}$ \emph{at the current particle position} exhibits high-strain and low-rotation.  Our analysis also shows how in flows where the centrifuge mechanism does not operate and the preferential sampling effect is absent, the non-local mechanism generates clustering without preferential concentration. 

\section{Theoretical Analysis}\label{Analysis}

\subsection{Derivation of the drift velocity conditioned on the local flow structure}

We consider the relative motion between two identical point particles, a `primary' particle and a `satellite' particle, the primary particle being the reference particle. The particles experience Stokes drag forces, but do not interact with each other through physical collisions or hydrodynamic interactions and are assumed to be at low enough concentration to not affect the turbulence (i.e., `one-way coupling'). The turbulence is assumed to be statistically stationary, homogeneous and isotropic, and gravitational settling is neglected. The particle-to-fluid density ratio is large, corresponding to a gas--solid system. Under these assumptions, the equation governing the relative motion of the two particles reduces to \citep{maxey83}
\begin{equation}
\dot{\boldsymbol{w}}^p(t)=(St\tau_{\eta})^{-1}\Big(\Delta\boldsymbol{u}(\boldsymbol{x}^p(t),\boldsymbol{r}^p(t),t)-\boldsymbol{w}^p(t)\Big),\label{eom}	
\end{equation}
where $\boldsymbol{r}^p(t),\boldsymbol{w}^p(t), \dot{\boldsymbol{w}}^p(t)$ are the particle pair relative separation, velocity and acceleration vectors, respectively, $\boldsymbol{x}^p(t)$ denotes the primary particle position and $\Delta\boldsymbol{u}(\boldsymbol{x}^p(t),\boldsymbol{r}^p(t),t)$ is the difference in the fluid velocity evaluated at the positions of the two particles.  Since we are interested in the case where $\boldsymbol{u}(\boldsymbol{x},t)$ is statistically homogeneous, we shall drop the $\boldsymbol{x}^p(t)$ argument when $\Delta\boldsymbol{u}$ appears in statistical expressions. 

For the system governed by (\ref{eom}) the exact equation governing the probability density function (PDF),
\[{p(\boldsymbol{r},\boldsymbol{w},t)\equiv\Big\langle\delta(\boldsymbol{r}^p(t)-\boldsymbol{r})\delta(\boldsymbol{w}^p(t)-\boldsymbol{w})\Big\rangle},\]
that describes the distribution of $\boldsymbol{r}^p(t),\boldsymbol{w}^p(t)$ in the phase-space $\boldsymbol{r},\boldsymbol{w}$  is
\begin{align}
\partial_t p=-\boldsymbol{\nabla_r\cdot}p\boldsymbol{w}+(St\tau_{\eta})^{-1}\boldsymbol{\nabla_w\cdot}p\boldsymbol{w}-(St\tau_{\eta})^{-1}\boldsymbol{\nabla_w\cdot}p\Big\langle\Delta\boldsymbol{u}(\boldsymbol{r}^p(t),t)\Big\rangle_{\boldsymbol{r},\boldsymbol{w}},\label{PDFeq}
\end{align}
where $\langle\cdot\rangle_{\boldsymbol{r},\boldsymbol{w}}$ denotes an ensemble average conditioned on ${\boldsymbol{r}^p(t)=\boldsymbol{r}}$ and ${\boldsymbol{w}^p(t)=\boldsymbol{w}}$.  
A commonly used statistical measure of particle clustering is the radial distribution function \citep[RDF,][]{mcquarrie}, which is defined as the ratio of the number of particle pairs at separation $\boldsymbol{r}$
to the number that would be expected if the particles were uniformly distributed. An exact equation for the statistically stationary RDF, $g(\boldsymbol{r})$, can be constructed by multiplying the stationary form of (\ref{PDFeq}) by $\boldsymbol{w}$ and then integrating over all $\boldsymbol{w}$ yielding
\begin{equation}
\boldsymbol{0}=g\Big\langle\Delta\boldsymbol{u}(\boldsymbol{r}^p(t),t)\Big\rangle_{\boldsymbol{r}}-St\tau_{\eta}\boldsymbol{S}^{p}_{2}\boldsymbol{\cdot\nabla_r}g-St\tau_{\eta}g\boldsymbol{\nabla_r\cdot}\boldsymbol{S}^{p}_{2},\label{RDFeq}
\end{equation}
where 
\begin{align}
g(\boldsymbol{r})=\frac{N(N-1)}{n^2 V}\int_{\boldsymbol{w}} p(\boldsymbol{r},\boldsymbol{w})\,d\boldsymbol{w},	
\end{align}
$N$ is the total number of particles lying within the control volume $V$, $n\equiv N/V$ is the number density of particles, and ${\boldsymbol{S}^{p}_{2}(\boldsymbol{r})\equiv\langle\boldsymbol{w}^p(t)\boldsymbol{w}^p(t)\rangle_{\boldsymbol{r}}}$ is the second-order particle velocity structure function.

Equation (\ref{RDFeq}) is exact, but unclosed, due to the term $\langle\Delta\boldsymbol{u}(\boldsymbol{r}^p(t),t)\rangle_{\boldsymbol{r}}$ which describes the average of $\Delta\boldsymbol{u}$ experienced by particle pairs at the separation $\boldsymbol{r}^p(t)=\boldsymbol{r}$.  Even though $\langle\Delta\boldsymbol{u}(\boldsymbol{r},t)\rangle=\boldsymbol{0}$ for isotropic flows, $\langle\Delta\boldsymbol{u}(\boldsymbol{r}^p(t),t)\rangle_{\boldsymbol{r}}\neq\boldsymbol{0}$ for finite $St$ because the inertial particles preferentially sample the underlying field $\Delta\boldsymbol{u}(\boldsymbol{x},\boldsymbol{r},t)$. In the ZT, this term is closed by approximating $\Delta\boldsymbol{u}(\boldsymbol{x},\boldsymbol{r},t)$ as a spatio-temporally correlated Gaussian field and using the Furutsu-Novikov closure method \citep{zaichik07,bragg14b}. However, for the purposes of this analysis, it is not necessary to introduce any such closure approximations. Recall that the purpose of this analysis is to demonstrate that the non-local clustering mechanism, which dominates at ${{St=O(1)}}$, is consistent with the phenomenon of preferential concentration.  It is unnecessary to close $\langle\Delta\boldsymbol{u}(\boldsymbol{r}^p(t),t)\rangle_{\boldsymbol{r}}$ for two reasons. First, \cite{bragg14b} showed that at ${St=O(1)}$ the drift velocity contributions coming from $\langle\Delta\boldsymbol{u}(\boldsymbol{r}^p(t),t)\rangle_{\boldsymbol{r}}$ are negligible compared to the drift mechanism associated with $St\tau_{\eta}\boldsymbol{\nabla_r\cdot}\boldsymbol{S}^{p}_{2}$ in (\ref{RDFeq}).  The reason for this is that at ${St=O(1)}$ in the dissipation range of the turbulence, the particle relative velocities exhibit `caustics' \citep{wilkinson05,bec10a,salazar12a}, which implies they are much larger than the fluid relative velocities.  A consequence of this is that contributions to the particle relative motion coming from their local interaction with the turbulence (e.g. $\Delta\boldsymbol{u}(\boldsymbol{r}^p(t),t)$) become negligible.  Second, any contributions to the clustering mechanism associated with $\langle\Delta\boldsymbol{u}(\boldsymbol{r}^p(t),t)\rangle_{\boldsymbol{r}}$ are manifestly consistent with the phenomenon of preferential concentration since this term arises because of preferential sampling effects.

The challenge therefore is to show that the drift velocity ${St\tau_{\eta}\boldsymbol{\nabla_r\cdot}\boldsymbol{S}^{p}_{2}}$, which generates the clustering and is strongly non-local at ${{St=O(1)}}$, is consistent with the phenomenon of preferential concentration. The physical interpretation of ${St\tau_{\eta}\boldsymbol{\nabla_r\cdot}\boldsymbol{S}^{p}_{2}}$ has already been discussed in the introduction. We refer the reader to \cite{bragg14b} for a more detailed explanation.

Since we are interested in clustering in the dissipation range we take
\[\Delta\boldsymbol{u}(\boldsymbol{x}^p(t),\boldsymbol{r}^p(t),t)=\boldsymbol{\Gamma}(\boldsymbol{x}^p(t),t)\boldsymbol{\cdot}\boldsymbol{r}^p(t),\]
where $\boldsymbol{\Gamma}(\boldsymbol{x},t)\equiv\boldsymbol{\nabla_x u}(\boldsymbol{x},t)$ is the fluid velocity gradient tensor.
If the non-local clustering mechanism for particles with ${St=O(1)}$ is consistent with the phenomenon of preferential concentration, then it must be the case that contributions to $St\tau_{\eta}\boldsymbol{\nabla_r\cdot}\boldsymbol{S}^{p}_{2}$ are larger for particles that are (at present time, $t$) sitting in high-strain, low-rotation regions than they are for particles sitting in high-rotation, low-strain regions of the flow field.

In order to formally consider this possibility, we expand the phase-space description of the drift velocity as shown below
\begin{equation}
-St\tau_{\eta}\boldsymbol{\nabla_r\cdot}\boldsymbol{S}^p_2\equiv\int\limits_{\mathcal{Z}} \varrho(\mathcal{Z})\underbrace{\Bigg\{-St\tau_{\eta}\boldsymbol{\nabla_r\cdot}\Big\langle\boldsymbol{w}^p(t)\boldsymbol{w}^p(t)\Big\rangle_{\boldsymbol{r},\mathcal{Z}}\Bigg\}}_{\text{Drift velocity conditioned on $\mathcal{Z}^p(t,t)=\mathcal{Z}$}}\,d\mathcal{Z},\label{driftZdef}
\end{equation}
where $\mathcal{Z}$ is an independent phase-space variable, ${\varrho(\mathcal{Z})\equiv\langle\delta(\mathcal{Z}^p(t,t)-\mathcal{Z})\rangle}$,\newline${\mathcal{Z}^p(t^\prime,t^{\prime\prime})\equiv\boldsymbol{\mathcal{S}}(\boldsymbol{x}^p(t^\prime),t^\prime)\boldsymbol{:}\boldsymbol{\mathcal{S}}(\boldsymbol{x}^p(t^{\prime\prime}),t^{\prime\prime})-\boldsymbol{\mathcal{R}}(\boldsymbol{x}^p(t^\prime),t^\prime)\boldsymbol{:}\boldsymbol{\mathcal{R}}(\boldsymbol{x}^p(t^{\prime\prime}),t^{\prime\prime})}$, $\boldsymbol{\mathcal{S}}$ and $\boldsymbol{\mathcal{R}}$ are the fluid strain-rate and rotation-rate tensors, respectively, defined as $\boldsymbol{\mathcal{S}}\equiv(1/2)(\boldsymbol{\Gamma}+\boldsymbol{\Gamma}^\top)$, $\boldsymbol{\mathcal{R}}\equiv(1/2)(\boldsymbol{\Gamma}-\boldsymbol{\Gamma}^\top)$, and $\langle\cdot\rangle_{\boldsymbol{r},\mathcal{Z}}$ denotes an ensemble average conditioned on ${\boldsymbol{r}^p(t)=\boldsymbol{r}}$ and ${\mathcal{Z}^p(t,t)=\mathcal{Z}}$.  We have defined $\mathcal{Z}^p$ in terms of two time arguments $t^\prime$ and $t^{\prime\prime}$ since quantities appearing later in the analysis will involve the more general expression $\mathcal{Z}^p(t^\prime,t^{\prime\prime})$. Through (\ref{driftZdef}) we are able to consider how the non-local contribution to the inward drift velocity is related to the local properties of $\boldsymbol{\Gamma}$ at the particles \emph{current} location $\boldsymbol{x}^p(t)$, where the relevant properties of $\boldsymbol{\Gamma}(\boldsymbol{x}^p(t),t)$ are characterized by its invariant $\mathcal{Z}^p(t,t)$.

We shall now define what we mean by high-strain, low-rotation regions and high-rotation, low-strain regions.  A rigorous definition of these regions could be given in terms of the invariants and eigenvalues of $\boldsymbol{\Gamma}(\boldsymbol{x},t)$ \citep[e.g.][]{chong90,rouson01,salazar12b}, however for our purposes this is not necessary.  In the regime ${St\ll1}$, the single particle velocity $\boldsymbol{v}^p(t)$ can be approximated as being a field, i.e. ${\boldsymbol{v}^p(t)\approx\boldsymbol{v}(\boldsymbol{x}^p(t),t)}$ with a divergence ${\boldsymbol{\nabla_x\cdot}\boldsymbol{v}(\boldsymbol{x}^p(t),t)\approx-St\tau_\eta\mathcal{Z}^p(t,t)}$ implying particles will cluster in regions where ${\mathcal{Z}^p(t,t)>0}$ \citep{maxey87}, which are commonly referred to as high-strain, low-rotation regions. Numerical simulations have not only confirmed this for ${St\ll1}$, but have shown that inertial particles continue to preferentially concentrate in ${\mathcal{Z}^p(t,t)>0}$ regions even when ${{St=O(1)}}$. Since it is the clustering behavior in these regions that is of interest, what we want to demonstrate is that for ${{St=O(1)}}$
\begin{align}
-St\tau_{\eta}\boldsymbol{\nabla_r\cdot}\Big\langle\boldsymbol{w}^p(t)\boldsymbol{w}^p(t)\Big\rangle_{\boldsymbol{r},\mathcal{Z}^{[+]}}<-St\tau_{\eta}\boldsymbol{\nabla_r\cdot}\Big\langle\boldsymbol{w}^p(t)\boldsymbol{w}^p(t)\Big\rangle_{\boldsymbol{r},\mathcal{Z}^{[-]}},
\end{align}
where
\[-St\tau_{\eta}\boldsymbol{\nabla_r\cdot}\Big\langle\boldsymbol{w}^p(t)\boldsymbol{w}^p(t)\Big\rangle_{\boldsymbol{r},\mathcal{Z}^{[+]}}\equiv\int\limits_{0}^{+\infty} \varrho(\mathcal{Z})\Bigg\{-St\tau_{\eta}\boldsymbol{\nabla_r\cdot}\Big\langle\boldsymbol{w}^p(t)\boldsymbol{w}^p(t)\Big\rangle_{\boldsymbol{r},\mathcal{Z}}\Bigg\}\,d\mathcal{Z},\]
and
\[-St\tau_{\eta}\boldsymbol{\nabla_r\cdot}\Big\langle\boldsymbol{w}^p(t)\boldsymbol{w}^p(t)\Big\rangle_{\boldsymbol{r},\mathcal{Z}^{[-]}}\equiv\int\limits_{-\infty}^{0} \varrho(\mathcal{Z})\Bigg\{-St\tau_{\eta}\boldsymbol{\nabla_r\cdot}\Big\langle\boldsymbol{w}^p(t)\boldsymbol{w}^p(t)\Big\rangle_{\boldsymbol{r},\mathcal{Z}}\Bigg\}\,d\mathcal{Z}.\]
Note that by definition
\[-St\tau_{\eta}\boldsymbol{\nabla_r\cdot}\boldsymbol{S}^p_2\equiv-St\tau_{\eta}\boldsymbol{\nabla_r\cdot}\Big\langle\boldsymbol{w}^p(t)\boldsymbol{w}^p(t)\Big\rangle_{\boldsymbol{r},\mathcal{Z}^{[+]}}-St\tau_{\eta}\boldsymbol{\nabla_r\cdot}\Big\langle\boldsymbol{w}^p(t)\boldsymbol{w}^p(t)\Big\rangle_{\boldsymbol{r},\mathcal{Z}^{[-]}}.\]

The formal solution for $\boldsymbol{w}^p(t)$ for particles in the dissipation range governed by (\ref{eom}) is (ignoring initial conditions since we are interested in the statistically stationary state)
\begin{align}
\boldsymbol{w}^p(t)=(St\tau_\eta)^{-1}\int\limits_{0}^{t} e^{-(t-t^\prime)/\tau_p}\boldsymbol{\Gamma}(\boldsymbol{x}^p(t^\prime),t^\prime)\boldsymbol{\cdot}\boldsymbol{r}^p(t^\prime)\,d t^\prime,
\end{align}
and using this we obtain
\begin{equation}
-St\tau_\eta\boldsymbol{\nabla_r\cdot}\Big\langle\boldsymbol{w}^p(t)\boldsymbol{w}^p(t)\Big\rangle_{\boldsymbol{r},\mathcal{Z}}=-(St\tau_\eta)^{-1}\int\limits_0^t\int\limits_0^t e^{-(2t-t^\prime-t^{\prime\prime})/\tau_p}\boldsymbol{\nabla_r\cdot}\boldsymbol{Q}\,dt^{\prime} dt^{\prime\prime},\label{wwZ}
\end{equation}
where the components of $\boldsymbol{Q}$ are 
\begin{equation}
{Q}_{ij}(\boldsymbol{r},\mathcal{Z},t,t^\prime,t^{\prime\prime})\equiv\Big\langle{\Gamma}_{im}(\boldsymbol{x}^p(t^\prime),t^\prime){r}_m^p(t^\prime){\Gamma}_{jn}(\boldsymbol{x}^p(t^{\prime\prime}),t^{\prime\prime}){r}_n^p(t^{\prime\prime})\Big\rangle_{\boldsymbol{r},\mathcal{Z}}.\label{Q}	
\end{equation}
We now seek a simplification of (\ref{Q}) that is valid in the regime of interest, namely ${St=O(1)}$ and $r\ll\eta$, where $\eta$ is the Kolmogorov length scale. Since we are interested in ${St=O(1)}$, neither $St$ nor $1/St$ can be treated as a small parameter. However in this regime, the particle velocity dynamics are strongly non-local, implying ${\boldsymbol{w}^p\gg\Delta\boldsymbol{u}}$, which is associated with the formation of caustics in the particle velocities \citep{wilkinson05,bec10a,salazar12a}.  This suggests that we can define the small parameter ${\mu\equiv\vert\langle\Delta\boldsymbol{u}(\boldsymbol{r},t)\Delta\boldsymbol{u}(\boldsymbol{r},t)\rangle\vert/\vert\boldsymbol{S}^p_2(\boldsymbol{r})\vert}$, which in the caustic regions satisfies $\mu\ll1$.  We therefore introduce the perturbation expansion  
\begin{equation}
\boldsymbol{Q}=\boldsymbol{Q}^{[0]}+\mu\boldsymbol{Q}^{[1]}+\mu^2\boldsymbol{Q}^{[2]}+\cdot\cdot\cdot,\label{QPT}	
\end{equation}
where the components of $\boldsymbol{Q}^{[0]}$ are   
\begin{equation}
{Q}_{ij}^{[0]}=\Big\langle{r}_{m}^{p}(t^\prime){r}_n^p(t^{\prime\prime})\Big\rangle_{\boldsymbol{r}}\Big\langle{\Gamma}_{im}(\boldsymbol{x}^p(t^\prime),t^\prime){\Gamma}_{jn}(\boldsymbol{x}^p(t^{\prime\prime}),t^{\prime\prime})\Big\rangle_{\mathcal{Z}}.\label{Q0}	
\end{equation}
This expression for $\boldsymbol{Q}^{[0]}$ follows from the fact that in the limit ${\mu\to0}$ the particle motion approaches ballistic motion, implying that the correlation between $\boldsymbol{r}^p$ and $\boldsymbol{\Gamma}$ vanishes, and the particles uniformly sample $\boldsymbol{\Gamma}$. In the regime ${\mu\ll1}$, to leading order ${\boldsymbol{Q}\approx\boldsymbol{Q}^{[0]}+O(\mu)}$, which can be used in (\ref{wwZ}) to construct an expression for ${-St\tau_\eta\boldsymbol{\nabla_r\cdot}\langle\boldsymbol{w}^p(t)\boldsymbol{w}^p(t)\rangle_{\boldsymbol{r},\mathcal{Z}}}$.

The term $\langle\boldsymbol{r}^p(t^\prime)\boldsymbol{r}^p(t^{\prime\prime})\rangle_{\boldsymbol{r}}$ in (\ref{Q0}) is related to the average growth of the particle pair separation $\boldsymbol{r}^p$ backward in time (noting that $t^{\prime}\leq t$ and $t^{\prime\prime}\leq t$), evaluated along pair trajectories that satisfy $\boldsymbol{r}^p(t)=\boldsymbol{r}$.  For $t^{\prime}=t^{\prime\prime}$ the quantity is the backward-in-time mean square separation of the particle pair.  The quantity may be expressed as
\begin{equation}
\Big\langle\boldsymbol{r}^p(t^\prime)\boldsymbol{r}^p(t^{\prime\prime})\Big\rangle_{\boldsymbol{r}}\equiv\boldsymbol{rr}+\boldsymbol{\Theta},\label{RRapx}
\end{equation}
where $\boldsymbol{\Theta}(\boldsymbol{r},t,t^{\prime},t^{\prime\prime})$ is the backward-in-time separation tensor, satisfying ${\boldsymbol{\Theta}(\boldsymbol{r},t,t,t)=\boldsymbol{0}}$.  In the regime ${\mu\ll1}$, using the formal solution for $\boldsymbol{r}^p$ for particles governed by (\ref{eom}), we obtain the approximation \citep[see][]{bragg14a}
\begin{equation}
\boldsymbol{\Theta}=(St\tau_\eta)^2\boldsymbol{S}^p_2\Big[1-e^{(t-t^\prime)/\tau_p}-e^{(t-t^{\prime\prime})/\tau_p}+e^{(2t-t^\prime-t^{\prime\prime})/\tau_p}\Big]+O(\mu).\label{ThetaDef}
\end{equation}
The result in (\ref{ThetaDef}) is only valid for $\max[t-t^\prime,t-t^{\prime\prime}]\leq O(\tau_p)$ since for larger time separations the process transitions into a different regime \citep{bragg14a}.  However, particles only retain a memory of their interaction with the turbulence up to times $O(t-\tau_p)$ in their path-history, making the approximation in (\ref{ThetaDef}) sufficient to capture the dominant contribution to the drift velocity. 

Substituting (\ref{RRapx}) into (\ref{Q0}), and taking the divergence yields
\begin{align}
\begin{split}
\nabla_{r_j}Q_{ij}=&\quad\,\nabla_{r_j}[r_m r_n]\Big\langle{\Gamma}_{im}(\boldsymbol{x}^p(t^\prime),t^\prime){\Gamma}_{jn}(\boldsymbol{x}^p(t^{\prime\prime}),t^{\prime\prime})\Big\rangle_{\mathcal{Z}}\\
&+\nabla_{r_j}[\Theta_{mn}]\Big\langle{\Gamma}_{im}(\boldsymbol{x}^p(t^\prime),t^\prime){\Gamma}_{jn}(\boldsymbol{x}^p(t^{\prime\prime}),t^{\prime\prime})\Big\rangle_{\mathcal{Z}}+O(\mu).
\end{split}
\end{align}
Using (\ref{ThetaDef}) in this expression, and substituting this into (\ref{wwZ}) and simplifying yields
\begin{equation}
\begin{split}
-St\tau_\eta\boldsymbol{\nabla_r\cdot}\Big\langle\boldsymbol{w}^p(t)\boldsymbol{w}^p(t)\Big\rangle_{\boldsymbol{r},\mathcal{Z}^{[+,-]}}&=\underbrace{-(St\tau_\eta)^{-1}\frac{\boldsymbol{r}}{3}\int\limits_{0}^t\int\limits_{0}^t e^{-(2t-t^\prime-t^{\prime\prime})/\tau_p}\Big\langle\mathcal{Z}^p(t^\prime,t^{\prime\prime})\Big\rangle_{\mathcal{Z}^{[+,-]}}\,dt^{\prime} dt^{\prime\prime}}_{\mycirc{\textbf{1}}}\\
&\,\,\qquad\underbrace{-(St\tau_\eta )^{-1}\frac{\boldsymbol{r}}{3}\int\limits_{0}^t\int\limits_{0}^t r^{-1}\nabla_r\widetilde{\Theta}_{\parallel}\Big\langle\mathcal{Y}^p(t^\prime,t^{\prime\prime})\Big\rangle_{\mathcal{Z}^{[+,-]}}\,dt^{\prime} dt^{\prime\prime}}_{\mycirc{\textbf{2}}},
\label{DriftZ}
\end{split}
\end{equation}
where ${\mathcal{Y}^p(t^\prime,t^{\prime\prime})\equiv\boldsymbol{\mathcal{S}}(\boldsymbol{x}^p(t^\prime),t^\prime)\boldsymbol{:}\boldsymbol{\mathcal{S}}(\boldsymbol{x}^p(t^{\prime\prime}),t^{\prime\prime})+\boldsymbol{\mathcal{R}}(\boldsymbol{x}^p(t^\prime),t^\prime)\boldsymbol{:}\boldsymbol{\mathcal{R}}(\boldsymbol{x}^p(t^{\prime\prime}),t^{\prime\prime})}$,\newline${\widetilde{\Theta}_{\parallel}(r,t,t^\prime,t^{\prime\prime})\equiv r^{-2}\boldsymbol{rr:\Theta}(\boldsymbol{r},-t,-t^\prime,-t^{\prime\prime})}$, and we have introduced the approximation in $\boldsymbol{\Theta}$ that ${\boldsymbol{S}^p_{2}\approx S^p_{2\parallel}\mathbf{I}}$, which is valid for ${St\geq O(1)}$ \citep{wwz00}.  The notation $\mathcal{Z}^{[+,-]}$ in (\ref{DriftZ}) denotes that the conditionality is either to be taken as $\mathcal{Z}^{[+]}$ or $\mathcal{Z}^{[-]}$.
%
For an isotropic system, the drift velocity between the particles is given by the projection of (\ref{DriftZ}) along $r^{-1}\boldsymbol{r}$ such that the drift velocity only depends on $r=\vert\boldsymbol{r}\vert$ and not the vector $\boldsymbol{r}$.  Consequently, the sign of $\boldsymbol{r}$ in (\ref{DriftZ}) is irrelevant in the following discussion.

The physical interpretation of terms \mycirc{\textbf{1}} and \mycirc{\textbf{2}} may be explained as follows. Suppose that during the particle timescale, $\tau_p$, the pair separation $\boldsymbol{r}^p$ remains almost constant. In this case term~\mycirc{\textbf{2}} would vanish (because $\widetilde{\Theta}_{\parallel}$ and $\nabla_r\widetilde{\Theta}_{\parallel}$ are zero if $\boldsymbol{r}^p$ is constant, see (\ref{RRapx})) but term~\mycirc{\textbf{1}} would be finite and describes a drift velocity arising because the primary particle preferentially samples $\boldsymbol{\Gamma}$ along its trajectory. This scenario occurs in the regime ${St\ll1}$, as $\tau_p$ is so small that $\boldsymbol{r}^p$ remains nearly constant over the timescale of the particles. Consequently, term~\mycirc{\textbf{1}} $\gg$ term~\mycirc{\textbf{2}} and the inward drift arises from the biased sampling of $\boldsymbol{\Gamma}$ along the path-history of the particle for times up to $O(t-\tau_p)$. In this regime, the sum of term~\mycirc{\textbf{1}} for $\mathcal{Z}^{[+]}$ and $\mathcal{Z}^{[-]}$ yields the CT result \citep{chun05,bragg14b}
\[-St\tau_{\eta}\boldsymbol{\nabla_r\cdot}\boldsymbol{S}^{p}_{2}=-St\tau_\eta\frac{\boldsymbol{r}}{3}\Big\langle\mathcal{Z}^p(t,t)\Big\rangle.\]

For ${{St=O(1)}}$, in contrast, the pair separation $\boldsymbol{r}^p$ changes significantly over the timescale of the particles.  This change in their relative separation affects the fluid velocity differences the particles experience along their path-history since ${\Delta\boldsymbol{u}(\boldsymbol{x}^p(t),\boldsymbol{r}^p(t),t)\propto\boldsymbol{r}^p(t)}$ in the dissipation range, and this gives rise to the path-history symmetry breaking effect described earlier. The time dependent relative separation between the primary and satellite particles is described by $\boldsymbol{\Theta}$ (see (\ref{RRapx})), which appears in term~\mycirc{\textbf{2}} through ${\widetilde{\Theta}_\parallel}$.  The growth of $\boldsymbol{\Theta}$ backward in time ensures that particle pairs arriving at $\boldsymbol{r},t$ from larger separations dominate the behavior of $\langle\boldsymbol{w}^p(t)\boldsymbol{w}^p(t)\rangle_{\boldsymbol{r},\mathcal{Z}^{[+,-]}}$ relative to particle pairs arriving at $\boldsymbol{r},t$ from smaller separations.  An important point is that although ${St= O(1)}$ and ${\Delta\boldsymbol{u}\propto\boldsymbol{r}}$ gives rise to the operation of the non-local, path-history symmetry breaking mechanism, the strength of the resulting inward drift in the dissipation range is influenced by the way the particles have interacted with $\boldsymbol{\mathcal{S}}$ and $\boldsymbol{\mathcal{R}}$ along their path-history. In term~\mycirc{\textbf{2}}, this dependence is described by $\langle\mathcal{Y}^p(t^\prime,t^{\prime\prime})\rangle_{\mathcal{Z}^{[+,-]}}$. We will show that it is the fact that inertial particles interact with the fields $\boldsymbol{\mathcal{S}}$ and $\boldsymbol{\mathcal{R}}$ differently that allows the non-local clustering mechanism to generate stronger inward drift velocities into high-strain regions of the turbulence.

\subsection{Analysis of the drift velocity for finite flow time scales}\label{Anal}

Let us now consider how preferential sampling of $\boldsymbol{\Gamma}$ along the path-history of the particles influences terms~\mycirc{\textbf{1}} and \mycirc{\textbf{2}}, and in particular, how this affects the strength of the drift into $\mathcal{Z}^{[+]}$ and $\mathcal{Z}^{[-]}$ regions.  In what follows, we assume that the autocorrelations associated with $\boldsymbol{\mathcal{S}}(\boldsymbol{x}^p(t),t)$ and $\boldsymbol{\mathcal{R}}(\boldsymbol{x}^p(t),t)$ are positive (at least for ${\max[t-t^\prime,t-t^{\prime\prime}]\leq O(\tau_p)}$).  This is known to be true for ${St=0}$ particles \citep[e.g.,][]{gir90} and it seems reasonable to assume that this is also true for ${St>0}$ particles (in \S\ref{DNStest} we will show using DNS data that this is indeed the case).

Term~\mycirc{\textbf{1}} in (\ref{DriftZ}) contains \[\Big\langle\mathcal{Z}^p(t^\prime,t^{\prime\prime})\Big\rangle_{\mathcal{Z}^{[+,-]}}\equiv \Big\langle\boldsymbol{\mathcal{S}}(\boldsymbol{x}^p(t^\prime),t^\prime)\boldsymbol{:}\boldsymbol{\mathcal{S}}(\boldsymbol{x}^p(t^{\prime\prime}),t^{\prime\prime})-\boldsymbol{\mathcal{R}}(\boldsymbol{x}^p(t^\prime),t^\prime)\boldsymbol{:}\boldsymbol{\mathcal{R}}(\boldsymbol{x}^p(t^{\prime\prime}),t^{\prime\prime})\Big\rangle_{\mathcal{Z}^{[+,-]}}.\]For $\mathcal{Z}^{[+]}$ the contribution involving $\boldsymbol{\mathcal{S}}$ dominates, and for $\mathcal{Z}^{[-]}$ the contribution involving $\boldsymbol{\mathcal{R}}$ dominates.  Consequently   
\begin{subequations}
\begin{align}
&\Big\langle\mathcal{Z}^p(t^\prime,t^{\prime\prime})\Big\rangle_{\mathcal{Z}^{[+]}}>0,\\
&\Big\langle\mathcal{Z}^p(t^\prime,t^{\prime\prime})\Big\rangle_{\mathcal{Z}^{[-]}}<0,	
\end{align}
\label{AmBPre}
\end{subequations}
which leads to the expectation that the mechanism associated with term~\mycirc{\textbf{1}} causes the particles to drift out of $\mathcal{Z}^{[-]}$ regions and into $\mathcal{Z}^{[+]}$ regions. Therefore term~\mycirc{\textbf{1}} is consistent with preferential concentration.

An implicit assumption made in (\ref{AmBPre}) is that the expression ${\langle\mathcal{Z}^p(t^\prime,t^{\prime\prime})\rangle_{\mathcal{Z}^{[+,-]}}}$ remains of the same sign for ${\max[t-t^\prime,t-t^{\prime\prime}]\leq O(\tau_p)}$.  That is, if a particle is in a $\mathcal{Z}^{[+]}$ region at time $t$, then it has, on average, been in a $\mathcal{Z}^{[+]}$ region for as long as it can remember its interaction with the turbulence along its path-history, and similarly for $\mathcal{Z}^{[-]}$.  As an estimate, this is satisfied provided that the Lagrangian timescales of $\boldsymbol{\mathcal{S}}(\boldsymbol{x}^p(t),t)$ and $\boldsymbol{\mathcal{R}}(\boldsymbol{x}^p(t),t)$, $\tau_{\mathcal{S}}$ and $\tau_{\mathcal{R}}$, respectively, satisfy ${\tau_{\mathcal{S}}\geq O(\tau_p)}$ and ${\tau_{\mathcal{R}}\geq O(\tau_p)}$.  DNS data \citep{ireland14} shows that this condition is satisfied for ${St< O(10)}$.  Another, associated reason why the conditions ${\tau_{\mathcal{S}}\geq O(\tau_p)}$ and ${\tau_{\mathcal{R}}\geq O(\tau_p)}$ must be satisfied is that if ${\tau_{\mathcal{S}}\ll\tau_p}$ and ${\tau_{\mathcal{R}}\ll\tau_p}$ then the dominant contribution to the drift velocity would be from the particles' interaction with $\boldsymbol{\Gamma}$ in regions that are uncorrelated with $\boldsymbol{\Gamma}$ at its current position.  In this situation the bias in the drift velocity for given values of $\mathcal{Z}$ would necessarily vanish, as would the preferential concentration, a scenario that we shall return to later.

Term~\mycirc{\textbf{2}} contains  \[\Big\langle\mathcal{Y}^p(t^\prime,t^{\prime\prime})\Big\rangle_{\mathcal{Z}^{[+,-]}}\equiv \Big\langle\boldsymbol{\mathcal{S}}(\boldsymbol{x}^p(t^\prime),t^\prime)\boldsymbol{:}\boldsymbol{\mathcal{S}}(\boldsymbol{x}^p(t^{\prime\prime}),t^{\prime\prime})+\boldsymbol{\mathcal{R}}(\boldsymbol{x}^p(t^\prime),t^\prime)\boldsymbol{:}\boldsymbol{\mathcal{R}}(\boldsymbol{x}^p(t^{\prime\prime}),t^{\prime\prime})\Big\rangle_{\mathcal{Z}^{[+,-]}}.\]Similar to ${\langle\mathcal{Z}^p(t^\prime,t^{\prime\prime})\rangle_{\mathcal{Z}^{[+,-]}}}$, the contribution involving $\boldsymbol{\mathcal{S}}$ dominates ${\langle\mathcal{Y}^p(t^\prime,t^{\prime\prime})\rangle_{\mathcal{Z}^{[+]}}}$, and the contribution involving $\boldsymbol{\mathcal{R}}$ dominates ${\langle\mathcal{Y}^p(t^\prime,t^{\prime\prime})\rangle_{\mathcal{Z}^{[-]}}}$.  Unlike ${\langle\mathcal{Z}^p(t^\prime,t^{\prime\prime})\rangle_{\mathcal{Z}^{[+,-]}}}$ however, ${\langle\mathcal{Y}^p(t^\prime,t^{\prime\prime})\rangle_{\mathcal{Z}^{[+,-]}}}$ is of the same sign for $\mathcal{Z}^{[+]}$ and $\mathcal{Z}^{[-]}$ and therefore the non-local clustering mechanism associated with term~\mycirc{\textbf{2}} causes the particles to drift into both $\mathcal{Z}^{[+]}$ and $\mathcal{Z}^{[-]}$ regions.   Nevertheless, since inertial particles with ${{St=O(1)}}$ preferentially sample strain-dominated regions of the flow and under sample rotation-dominated regions \citep[e.g.,][]{salazar12b} then we expect that
\begin{align}
\Big\langle\mathcal{Y}^p(t^\prime,t^{\prime\prime})\Big\rangle_{\mathcal{Z}^{[+]}}>\Big\langle\mathcal{Y}^p(t^\prime,t^{\prime\prime})\Big\rangle_{\mathcal{Z}^{[-]}}.\label{ApBPre}	
\end{align}
Consequently, based on the predicted inequalities in (\ref{AmBPre}) and (\ref{ApBPre}), and since ${\nabla_r\widetilde{\Theta}_{\parallel} >0}$, (\ref{DriftZ}) leads to the expectation that even for ${St=O(1)}$ when the clustering mechanism is strongly non-local, the particles continue to drift more strongly into $\mathcal{Z}^{[+]}$ regions, consistent with the phenomenon of preferential concentration.  The physical explanation is that because the inertial particles preferentially sample $\boldsymbol{\Gamma}$ along their trajectories, the fluid velocity differences driving the particles together are experienced by the particles to be greater in $\mathcal{Z}^{[+]}$ regions than in $\mathcal{Z}^{[-]}$ regions.   This preferential sampling is a consequence of the particles being centrifuged away from regions of strong rotation along their trajectories.  Nevertheless for ${{St=O(1)}}$, the centrifuge mechanism is not the primary cause of clustering, but rather it simply influences the clustering behavior through the way it causes particles to preferentially sample $\boldsymbol{\Gamma}$ along their path-histories.
A consequence of this is that ${St= O(1)}$ particles can cluster even in situations where the centrifuge mechanism does not operate, for which biased sampling and hence preferential concentration disappear.  We now consider this situation.

\subsection{Analysis of the drift velocity for white-noise flows}\label{Anal-white-noise}

In the limit where ${\tau_{\mathcal{S}}\to0}$ and ${\tau_{\mathcal{R}}\to0}$, such as in white-in-time random flows or for particles falling through turbulence in the limit of strong gravity \citep{bec14,gustavsson14}, the centrifuge mechanism does not operate and there is no preferential sampling of $\boldsymbol{\Gamma}$.  In this limit 
\begin{align}
\Big\langle\mathcal{Z}^p(t^\prime,t^{\prime\prime})\Big\rangle_{\mathcal{Z}^{[+,-]}}= 
\left\{ \begin{array}{ll}
\Big\langle\mathcal{Z}^p(t,t)\Big\rangle_{\mathcal{Z}^{[+,-]}},& t^{\prime\prime}=t^{\prime}=t,\\
\vspace{1mm}\\
Z(t^\prime)\delta(t^\prime-t^{\prime\prime}),& t^{\prime\prime}<t,\, t^{\prime}<t.\end{array} \right. \label{ApBwn}	
\end{align}
where
\begin{align}
Z(t^\prime)\equiv\int_{-\infty}^{+\infty}	\Big\langle\mathcal{Z}^p(t^\prime,t^{\prime\prime})\Big\rangle\, d t^{\prime\prime}.
\end{align}
The ${t^\prime<t}$, ${t^{\prime\prime}<t}$ result follows from the fact that in the limit ${\tau_{\mathcal{S}}\to0}$ and ${\tau_{\mathcal{R}}\to0}$, ${\langle\mathcal{Z}^p(t^\prime,t^{\prime\prime})\rangle_{\mathcal{Z}^{[+,-]}}=0}$ if $t^\prime\neq t^{\prime\prime}$, and since $\mathcal{Z}^p(t^\prime,t^{\prime})$ and $\mathcal{Z}^p(t,t)$ would be uncorrelated in this limit then ${\langle\mathcal{Z}^p(t^\prime,t^{\prime})\rangle_{\mathcal{Z}^{[+,-]}}=\langle\mathcal{Z}^p(t^\prime,t^{\prime})\rangle}$ for $t^\prime\neq t$.  Furthermore, in the statistically stationary state, ${\langle\mathcal{Z}^p(t^\prime,t^{\prime})\rangle=\langle\mathcal{Z}^p(t,t)\rangle=0\implies Z(t^\prime)=0}$ and
\[\Big\langle\mathcal{Z}^p(t,t)\Big\rangle_{\mathcal{Z}^{[+]}}+\Big\langle\mathcal{Z}^p(t,t)\Big\rangle_{\mathcal{Z}^{[-]}}=0,\]
such that the contribution from term~\mycirc{\textbf{1}} to the total drift velocity is zero. However, in the same limit 
\begin{align}
\Big\langle\mathcal{Y}^p(t^\prime,t^{\prime\prime})\Big\rangle_{\mathcal{Z}^{[+,-]}}= 
\left\{ \begin{array}{ll}
\Big\langle\mathcal{Y}^p(t,t)\Big\rangle_{\mathcal{Z}^{[+,-]}},& t^{\prime\prime}=t^{\prime}=t,\\
\vspace{1mm}\\
Y(t^\prime)\delta(t^\prime-t^{\prime\prime}),& t^{\prime\prime}<t,\, t^{\prime}<t.\end{array} \right.
\label{ApBwn}	
\end{align}
where 
\begin{align}
Y(t^\prime)\equiv\int_{-\infty}^{+\infty}	\Big\langle\mathcal{Y}^p(t^\prime,t^{\prime\prime})\Big\rangle\, d t^{\prime\prime}.
\end{align}
Unlike $\langle\mathcal{Z}^p(t^\prime,t^{\prime})\rangle$, ${\langle\mathcal{Y}^p(t^\prime,t^{\prime})\rangle\neq0\implies Y(t^\prime)\neq0}$ in the statistically stationary state.  The result in (\ref{ApBwn}) for ${t^{\prime\prime}=t^{\prime}=t}$ does not render term~\mycirc{\textbf{2}} dependent upon $\mathcal{Z}$ in the limit ${\tau_{\mathcal{S}}\to0}$ and ${\tau_{\mathcal{R}}\to0}$ because in the integrand of term~\mycirc{\textbf{2}}, $\langle\mathcal{Y}^p(t^\prime,t^{\prime\prime})\rangle_{\mathcal{Z}^{[+,-]}}$ is multiplied by $\nabla_r\widetilde{\Theta}_\parallel$, and for ${t^{\prime\prime}=t^{\prime}=t}$, ${\nabla_r\widetilde{\Theta}_\parallel=0}$.  Consequently, since the result in (\ref{ApBwn}) is independent of $\mathcal{Z}$ for $t^{\prime\prime}<t$, $t^{\prime}<t$, then in the limit ${\tau_{\mathcal{S}}\to0}$ and ${\tau_{\mathcal{R}}\to0}$, term~\mycirc{\textbf{2}} generates a finite contribution to ${-St\tau_\eta\boldsymbol{\nabla_r\cdot}\boldsymbol{S}^p_2}$ but without a bias towards any particular region in $\mathcal{Z}$ space.  That is, term~\mycirc{\textbf{2}} becomes independent of $\mathcal{Z}$ and hence the drift velocity is equally strong for all $\mathcal{Z}$.

\subsection{Comparison with Gustavsson \& Mehlig (2011)}

The result presented in \S\ref{Anal-white-noise} is consistent with the results derived by \cite{gustavsson11b}, who showed that in the limit ${\tau_{\mathcal{S}}\to0}$ and ${\tau_{\mathcal{R}}\to0}$, the particles cluster, but without the preferential concentration effect. As discussed in \cite{bragg14b}, the `non-ergodic' and `ergodic' clustering mechanisms described in \cite{gustavsson11b} are qualitatively similar to the clustering mechanisms described by $-St\tau_\eta\boldsymbol{\nabla_r\cdot}\boldsymbol{S}^p_2$ in the regime€ ${\tau_p\ll\min[\tau_{\mathcal{S}},\tau_{\mathcal{R}}]}$ and ${\tau_p\gg\max[\tau_{\mathcal{S}},\tau_{\mathcal{R}}]}$, respectively.  The terms non-ergodic/ergodic in this context refer to the presence/absence of preferential sampling effects.  Because of its perturbative construction, the clustering theory in \cite{gustavsson11b} was not able to describe the interplay between the non-ergodic and ergodic clustering mechanisms for ${{St=O(1)}}$ and ${\mathrm{Ku}\sim1}$ \citep[where $\mathrm{Ku}$ is the non-dimensional Kubo number, which is ${\sim1}$ for real turbulence, see][]{duncan05}, although their simulations indicate that both make a substantial contribution to the clustering in this regime.  The theoretical analysis in this paper demonstrates how these two effects contribute in this regime: The non-local, path-history symmetry breaking mechanism, which operates even in the purely ergodic limit, is nevertheless influenced by non-ergodic effects when ${{St=O(1)}}$ (and $\mathrm{Ku}\sim1$), and these non-ergodic effects are what allow the non-local clustering mechanism to generate preferential concentration in this regime.

\subsection{Validity of dissipation range scaling}

In closing this section we briefly comment on the validity of assuming dissipation range scaling for $\Delta\boldsymbol{u}(\boldsymbol{x},\boldsymbol{r},t)$.  Since $\boldsymbol{w}^p$ depends non-locally in time on $\Delta\boldsymbol{u}$, then by introducing ${\Delta\boldsymbol{u}(\boldsymbol{x}^p(t),\boldsymbol{r}^p(t),t)=\boldsymbol{\Gamma}(\boldsymbol{x}^p(t),t)\boldsymbol{\cdot}\boldsymbol{r}^p(t)}$ into (\ref{eom}) we are implicitly assuming in our analysis that this assumption is valid up to times $O(t-\tau_p)$ along the path-history of the particle pair.  The dissipation range scaling ${\Delta\boldsymbol{u}(\boldsymbol{x},\boldsymbol{r},t)\propto\boldsymbol{r}}$ is supposed to be only valid for ${r\ll\eta}$ \citep{pope}.  This could be problematic for our analysis of particles with ${{St=O(1)}}$ that may retain a memory of their interaction with the turbulence at ${\vert\boldsymbol{r}^p\vert>\eta}$.  However, the problem is mitigated by the fact that numerical studies of Navier-Stokes turbulence have shown that ${\Delta\boldsymbol{u}(\boldsymbol{x},\boldsymbol{r},t)\propto\boldsymbol{r}}$ up to ${r=O(10\eta)}$ \citep{ishihara09}. Thus, the effective dissipation length scale is an order of magnitude larger than the Kolmogorov length scale, likely due to nonlinear depletion effects in the turbulence \citep{frisch} that allow the viscosity to act on scales an order of magnitude larger than would be expected based on a scaling argument. ${{St=O(1)}}$ particles at ${\vert\boldsymbol{r}^p(t)\vert\ll\eta}$ have a temporal memory that is sufficiently small that interactions with the turbulence at separations greater than $ O(10\eta)$ in their path-history only weakly affect the drift, an argument supported by the results in \cite{ray13}.  

\section{Tests using DNS}\label{DNStest}

In this section, we use DNS data to test the arguments presented in the previous section.
%
In particular, we will calculate from DNS data
\[\Big\langle\mathcal{Z}^p(t^\prime,t^{\prime\prime})\Big\rangle_{\mathcal{Z}^{[+,-]}},\quad\Big\langle\mathcal{Y}^p(t^\prime,t^{\prime\prime})\Big\rangle_{\mathcal{Z}^{[+,-]}},\]
along with $S^p_{2\parallel}$ (required in ${\widetilde{\Theta}_{\parallel}}$), evaluate the integrals in (\ref{DriftZ}), and then compare the strength of the drift velocity in the $\mathcal{Z}^{[+]}$ and $\mathcal{Z}^{[-]}$ regions.  The DNS simulations were of statistically stationary, homogeneous and isotropic turbulence at ${Re_\lambda=224}$, where ${Re_\lambda}$ is the Taylor microscale Reynolds number.  Details on the DNS can be found in \cite{ireland13}.

In computing the results, we replace the lower integral limit of $0$ in (\ref{DriftZ}) with $t-\tau_p$.  The reason for doing this is that the closure approximation used for $\boldsymbol{\Theta}$ is only valid for ${\max[t-t^\prime,t-t^{\prime\prime}]\leq O(\tau_p)}$, and gives unphysical results at larger time separations.  However, the cut-off is sufficient to capture the dominant contribution to the drift velocity since, as discussed in \S\ref{Analysis}, the particles only retain a memory of their interaction with the turbulence for times up to $O(t-\tau_p)$ along their path-history.

To make the notation in what follows more succinct we define\[\boldsymbol{\mathfrak{d}}^{[+,-]}\equiv-St\tau_\eta\boldsymbol{\nabla_r\cdot}\Big\langle\boldsymbol{w}^p(t)\boldsymbol{w}^p(t)\Big\rangle_{\boldsymbol{r},\mathcal{Z}^{[+,-]}},\]and since we are considering isotropic turbulence it is only the parallel component of the drift velocity that is of interest, $\mathfrak{d}_{\parallel}^{[+,-]}=r^{-1}\boldsymbol{r\cdot}\boldsymbol{\mathfrak{d}}^{[+,-]}$.  We also use the additional subscript of either $1$ or $2$ (e.g. $\mathfrak{d}_{\parallel,1}^{[+]}$) to denote that we are considering the contribution to $\mathfrak{d}_{\parallel}^{[+,-]}$ coming from either term~$\mycirc{\textbf{1}}$ or term~$\mycirc{\textbf{2}}$ in (\ref{DriftZ}).

In order to examine whether the drift velocity contains a bias for $\mathcal{Z}^{[+]}$ or $\mathcal{Z}^{[-]}$ we consider 
\begin{align}
J_1&\equiv\mathfrak{d}_{\parallel,1}^{[-]}\Big/\mathfrak{d}_{\parallel,1}^{[+]},\\
J_2&\equiv\mathfrak{d}_{\parallel,2}^{[-]}\Big/\mathfrak{d}_{\parallel,2}^{[+]}.
\end{align}
Since the $\boldsymbol{r}$ dependence of the drift velocity contributions to terms~$\mycirc{\textbf{1}}$ and $\mycirc{\textbf{2}}$ are the same for $\mathcal{Z}^{[+]}$ and $\mathcal{Z}^{[-]}$, $J_1$ and $J_2$ are functions of $St$, but not of $\boldsymbol{r}$.  As discussed earlier, the sum of the $\mathcal{Z}^{[+]}$ and $\mathcal{Z}^{[-]}$ contributions from term~$\mycirc{\textbf{1}}$ is zero in the absence of preferential concentration, implying that in the absence of preferential concentration ${J_1=-1}$.  For term~$\mycirc{\textbf{2}}$ however, the absence of preferential concentration is denoted by ${J_2=1}$, since, as demonstrated in \S\ref{Anal}, in the absence of preferential concentration, term~$\mycirc{\textbf{2}}$ becomes independent of $\mathcal{Z}$.
%
%
{\begin{figure}
\centering
\subfloat[]
{\begin{overpic}
[trim = 0mm 60mm 20mm 60mm,scale=0.35,clip,tics=20]{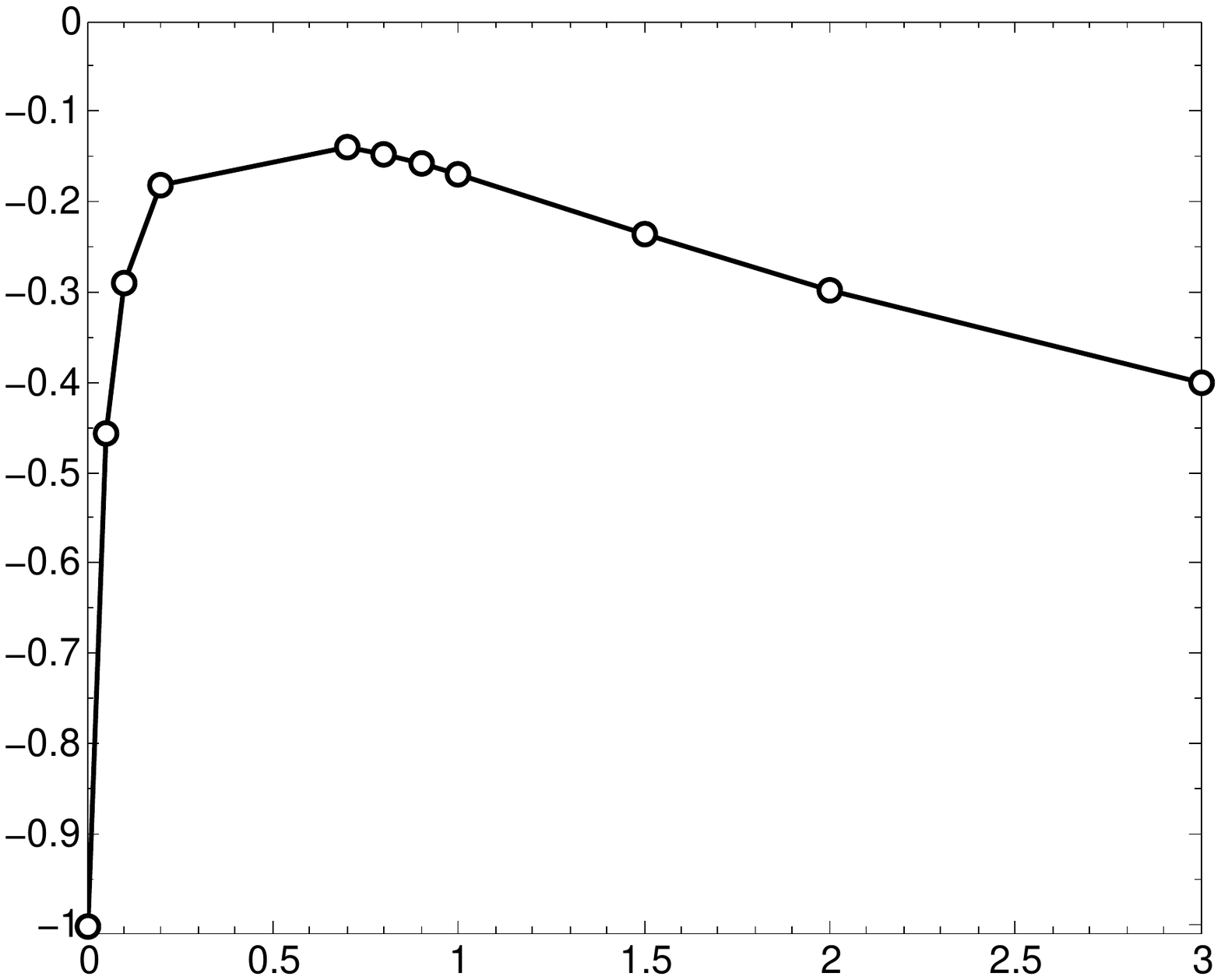}
\put(5,80){\rotatebox{0}{$J_1$}}
\put(107,0){$St$}
\end{overpic}}
\subfloat[]
{\begin{overpic}
[trim = 0mm 60mm 20mm 60mm,scale=0.35,clip,tics=20]{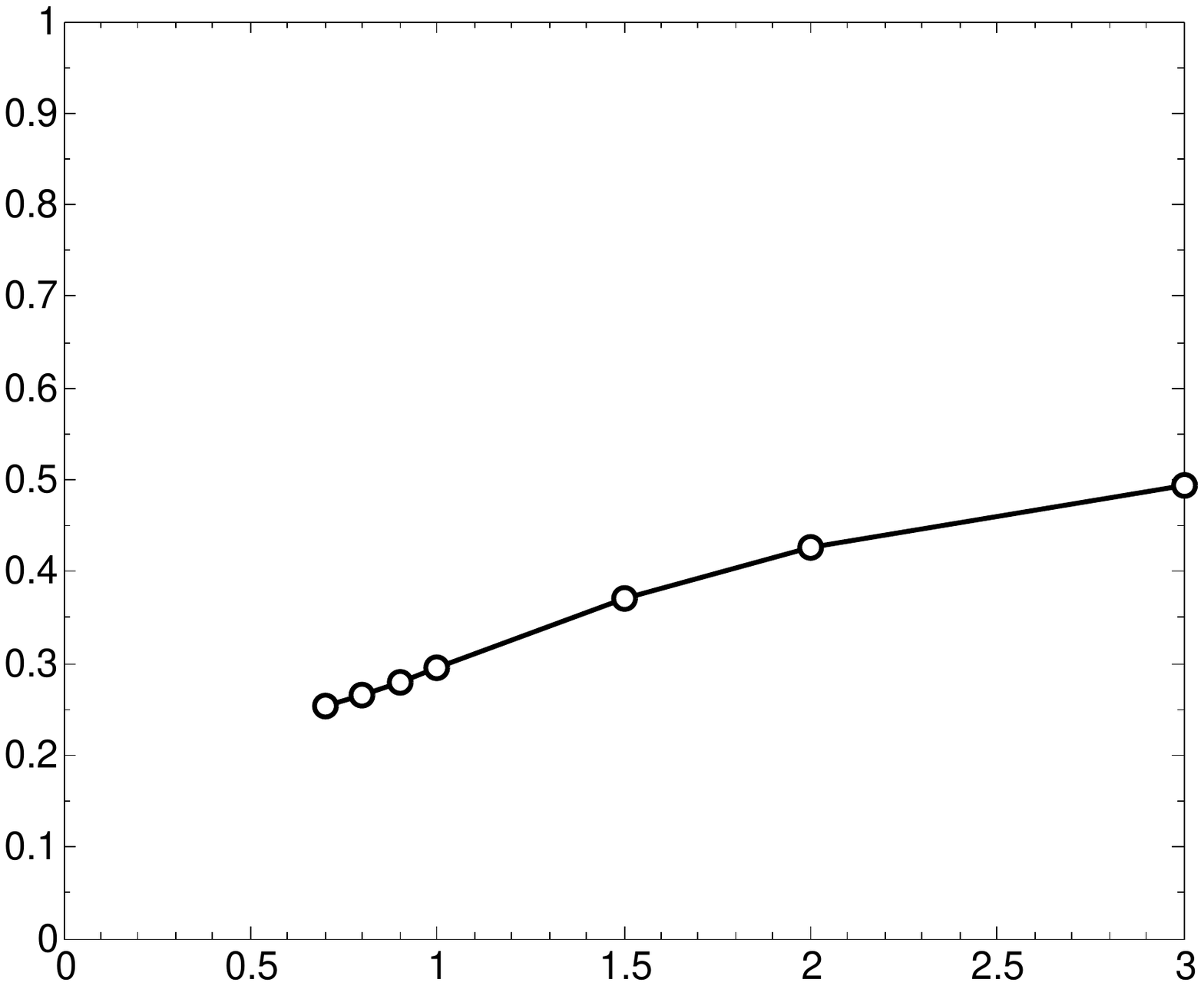}
\put(8,80){\rotatebox{0}{$J_2$}}
\put(107,0){$St$}
\end{overpic}}
\caption{Plots of (a) $J_1$ and (b) $J_2$ as a function of $St$.}
\label{TestDriftZ} 
\end{figure}
%
The results for $J_1$ and $J_2$ are shown in figure~\ref{TestDriftZ}.  For $St\geq 0.7$, $J_1$ and $J_2$ are evaluated using (\ref{DriftZ}); for $St\le 0.2$, we also evaluate $J_1$ and $J_2$ using the ${St\ll1}$ approximations  
\begin{align}
\mathfrak{d}_{\parallel,1}^{[+,-]}&\approx-\frac{St\tau_\eta}{3}r\Big\langle\mathcal{Z}^p(t,t)\Big\rangle_{\mathcal{Z}^{[+,-]}},\\
\mathfrak{d}_{\parallel,2}^{[-]}&=\mathfrak{d}_{\parallel,2}^{[+]}\approx0
\end{align}
(such that $J_2$ is undefined and is therefore not plotted for ${St\leq0.2}$).
The results in figure~\ref{TestDriftZ} show that for ${St>0}$, ${J_1>-1}$ and ${J_2<1}$, consistent with preferential concentration.   As $St$ is increased from $0$, $J_1$ initially increases, reaches a maximum at ${St\approx 0.7}$ and then begins to decrease again, consistent with known DNS data for the preferential concentration effect \citep[e.g., see figure 1 in][]{bec07}.  That ${-1\leq J_1<0}$ reflects the fact that the clustering mechanism associated with term~$\mycirc{\textbf{1}}$ causes particles to drift away from $\mathcal{Z}^{[-]}$ regions and drift into $\mathcal{Z}^{[+]}$, and is directly associated with the centrifuge mechanism acting along the particle trajectories.  In contrast, ${0\leq J_2<1}$ signifying that the mechanism associated with term~$\mycirc{\textbf{2}}$, the non-local clustering mechanism, causes particles to drift into both $\mathcal{Z}^{[+]}$ and $\mathcal{Z}^{[-]}$ regions.  However, because of the preferential sampling of the turbulence along the particle trajectories, the drift into $\mathcal{Z}^{[+]}$ regions is much stronger than that into $\mathcal{Z}^{[-]}$ regions.  These results clearly demonstrate that the non-local clustering mechanism that dominates $St\ge O(1)$ particles is fundamentally different to the traditional centrifuge mechanism, yet the particles continue to drift more strongly into $\mathcal{Z}^{[+]}$ regions than $\mathcal{Z}^{[-]}$ regions, consistent with the phenomenon of preferential concentration.



We now use the DNS data to test the key assumptions made in the theoretical analysis in \S\ref{Analysis}.
First, we made the assumption that for finite $St$, the Lagrangian autocorrelations of $\boldsymbol{\mathcal{S}}(\boldsymbol{x}^p(t),t)$ and $\boldsymbol{\mathcal{R}}(\boldsymbol{x}^p(t),t)$ are positive, although strictly speaking the analysis only requires that they remain positive for ${\max[t-t^\prime,t-t^{\prime\prime}]\leq O(\tau_p)}$. Figure~\ref{Strain_Rot_auto} shows the autocorrelations
\begin{align}
\Psi_{\mathcal{S}}(t,t^\prime)&\equiv\Big\langle\boldsymbol{\mathcal{S}}(\boldsymbol{x}^p(t),t)\boldsymbol{:}\boldsymbol{\mathcal{S}}(\boldsymbol{x}^p(t^{\prime}),t^{\prime})\Big\rangle\Big/\Big\langle\boldsymbol{\mathcal{S}}(\boldsymbol{x}^p(t),t)\boldsymbol{:}\boldsymbol{\mathcal{S}}(\boldsymbol{x}^p(t),t)\Big\rangle,\\
\Psi_{\mathcal{R}}(t,t^\prime)&\equiv\Big\langle\boldsymbol{\mathcal{R}}(\boldsymbol{x}^p(t),t)\boldsymbol{:}\boldsymbol{\mathcal{R}}(\boldsymbol{x}^p(t^{\prime}),t^{\prime})\Big\rangle\Big/\Big\langle\boldsymbol{\mathcal{R}}(\boldsymbol{x}^p(t),t)\boldsymbol{:}\boldsymbol{\mathcal{R}}(\boldsymbol{x}^p(t),t)\Big\rangle,	
\end{align}
for particles with ${{St=O(1)}}$.  The results clearly demonstrate that although $\Psi_{\mathcal{S}}(t,t^\prime)$ and $\Psi_{\mathcal{R}}(t,t^\prime)$ are affected by the particle inertia, they remain positive for ${t-t^\prime/\tau_p\leq O(1)}$, confirming the assumption.
{\begin{figure}
\centering
{\begin{overpic}
[trim = 20mm 65mm 20mm 0mm,scale=0.38,clip,tics=20]{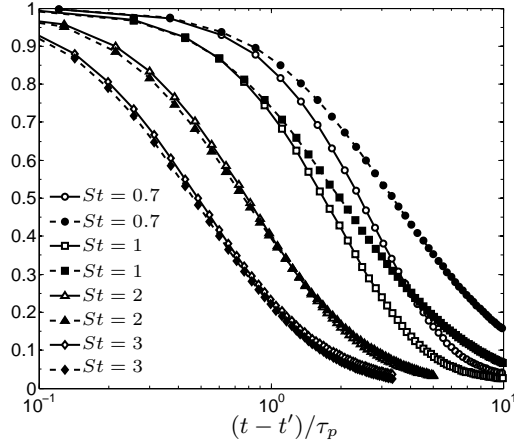}
\put(83,-5){$(t-t^{\prime})/\tau_p$}
\put(27,83){\scriptsize\text{$St=0.7$}}
\put(27,73){\scriptsize\text{$St=0.7$}}
\put(27,64){\scriptsize\text{$St=1$}}
\put(27,54){\scriptsize\text{$St=1$}}
\put(27,45){\scriptsize\text{$St=2$}}
\put(27,36){\scriptsize\text{$St=2$}}
\put(27,27){\scriptsize\text{$St=3$}}
\put(27,18){\scriptsize\text{$St=3$}}
\end{overpic}}
\caption{Plot of DNS data for $\Psi_{\mathcal{S}}$ (solid lines, open symbols) and $\Psi_{\mathcal{R}}$ (dashed lines, filled symbols) for various $St$.}
\label{Strain_Rot_auto} 
\end{figure}
%
A second assumption made in our analysis was that the autocovariances ${\langle\mathcal{Z}^p(t^\prime,t^{\prime\prime})\rangle_{\mathcal{Z}^{[+,-]}}}$ and ${\langle\mathcal{Y}^p(t^\prime,t^{\prime\prime})\rangle_{\mathcal{Z}^{[+,-]}}}$ remain of the same sign for ${\max[t-t^\prime,t-t^{\prime\prime}]\leq O(\tau_p)}$. Figure~\ref{Autocorr} plots the corresponding autocorrelations normalized by their value at ${t^\prime=t^{\prime\prime}=t}$, for ${St=1}$.  The results indicate that for sufficiently large $t-t^\prime,t-t^{\prime\prime}$ the autocovariances can change sign (see figure~\ref{Autocorr}a), reflecting the fact that a particle in a $\mathcal{Z}^{[-]}$ region at time $t$, for example, may have been in a $\mathcal{Z}^{[+]}$ region at some earlier time in its path-history.  However, the results confirm that for ${\max[t-t^\prime,t-t^{\prime\prime}]\leq O(\tau_p)}$ the autocovariances do not change sign, confirming the assumption in our analysis. 

A third and related assumption made in our analysis is that the timescales associated with ${\langle\mathcal{Z}^p(t^\prime,t^{\prime\prime})\rangle_{\mathcal{Z}^{[+,-]}}}$ and ${\langle\mathcal{Y}^p(t^\prime,t^{\prime\prime})\rangle_{\mathcal{Z}^{[+,-]}}}$ are ${\geq O(\tau_p)}$ which we estimated to be valid based on the known behavior of ${\tau_{\mathcal{S}}}$ and ${\tau_{\mathcal{R}}}$.  The results in figure~\ref{Autocorr} show that for ${\max[t-t^\prime,t-t^{\prime\prime}]\leq O(\tau_p)}$, the autocorrelations remain significant ($\gtrsim 0.5$) demonstrating that over the memory timescale of the particle, the strain and rotation fields remain significantly correlated along the particle path history.

These results validate the key assumptions made in our analysis.
\begin{figure}
\vspace{-10mm}
\centering
\subfloat[]
{\begin{overpic}
[trim = 0mm 55mm 0mm 50mm,scale=0.35,clip,tics=20]{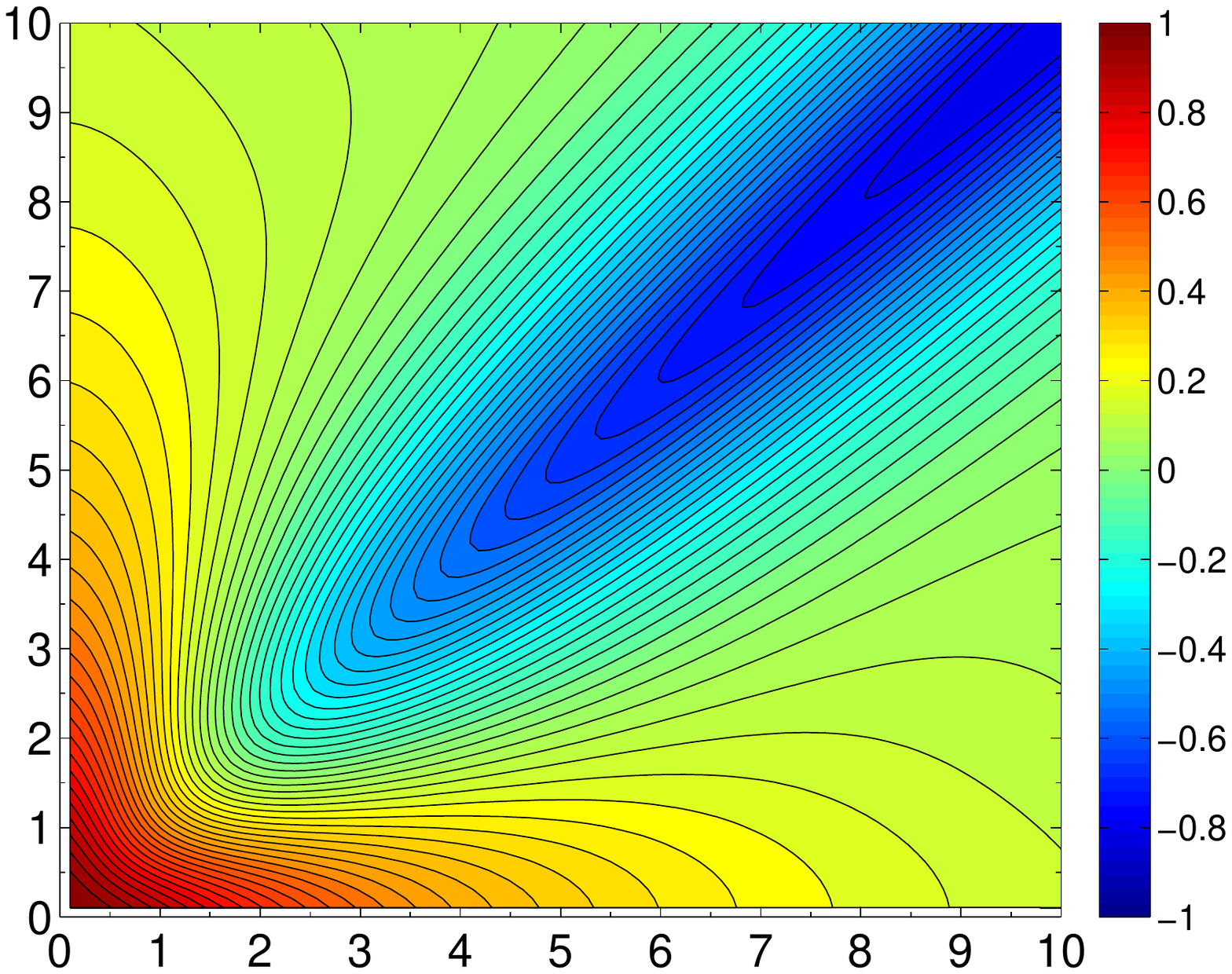}
\put(10,70){\rotatebox{90}{$(t-t^\prime)/\tau_p$}}
\put(87,3){$(t-t^{\prime\prime})/\tau_p$}
\end{overpic}}
\subfloat[]
{\begin{overpic}
[trim = 20mm 55mm 0mm 50mm,scale=0.35,clip,tics=20]{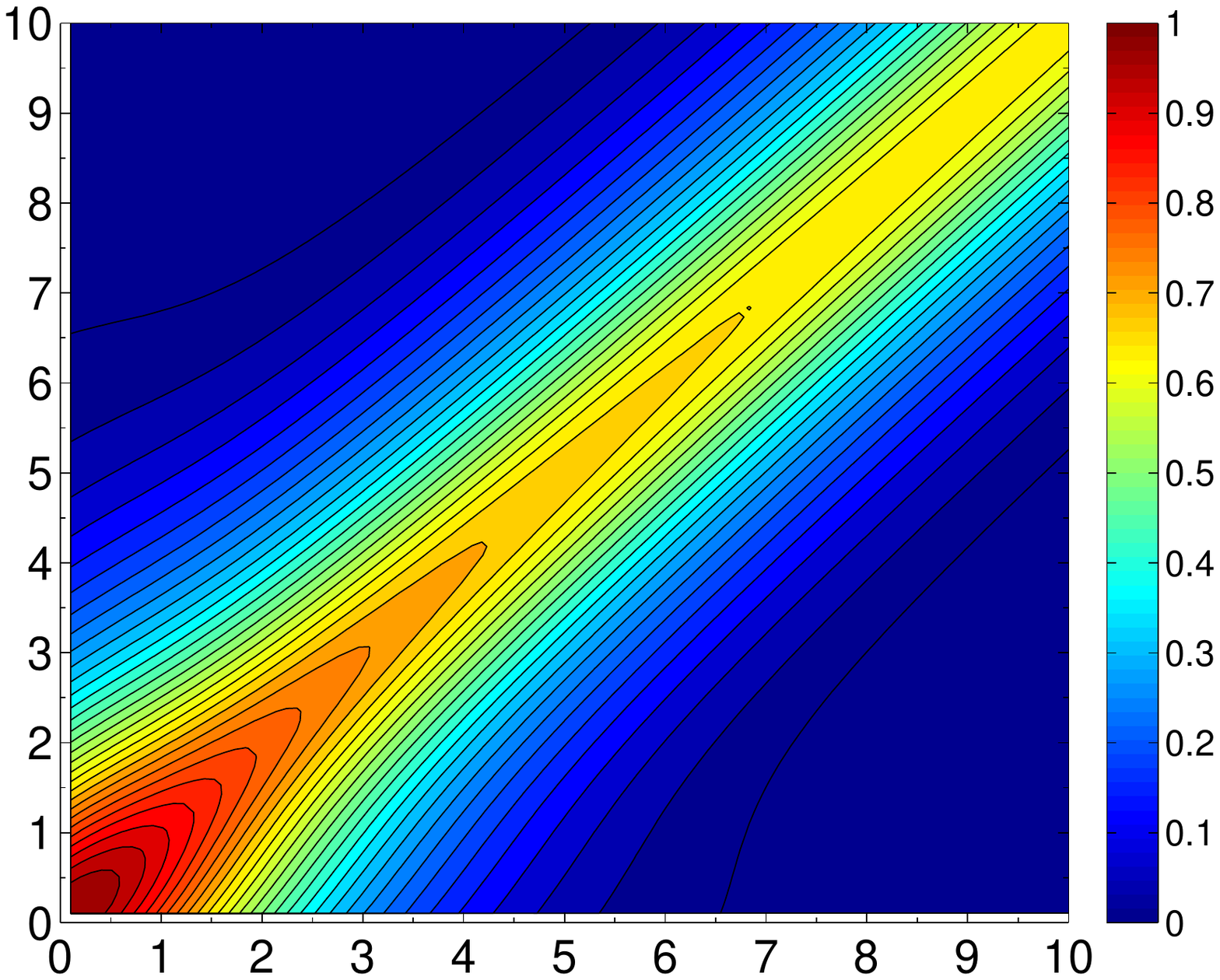}
\put(68,3){$(t-t^{\prime\prime})/\tau_p$}
\end{overpic}}\\
\vspace{-10mm}
\subfloat[]
{\begin{overpic}
[trim = 0mm 55mm 0mm 50mm,scale=0.35,clip,tics=20]{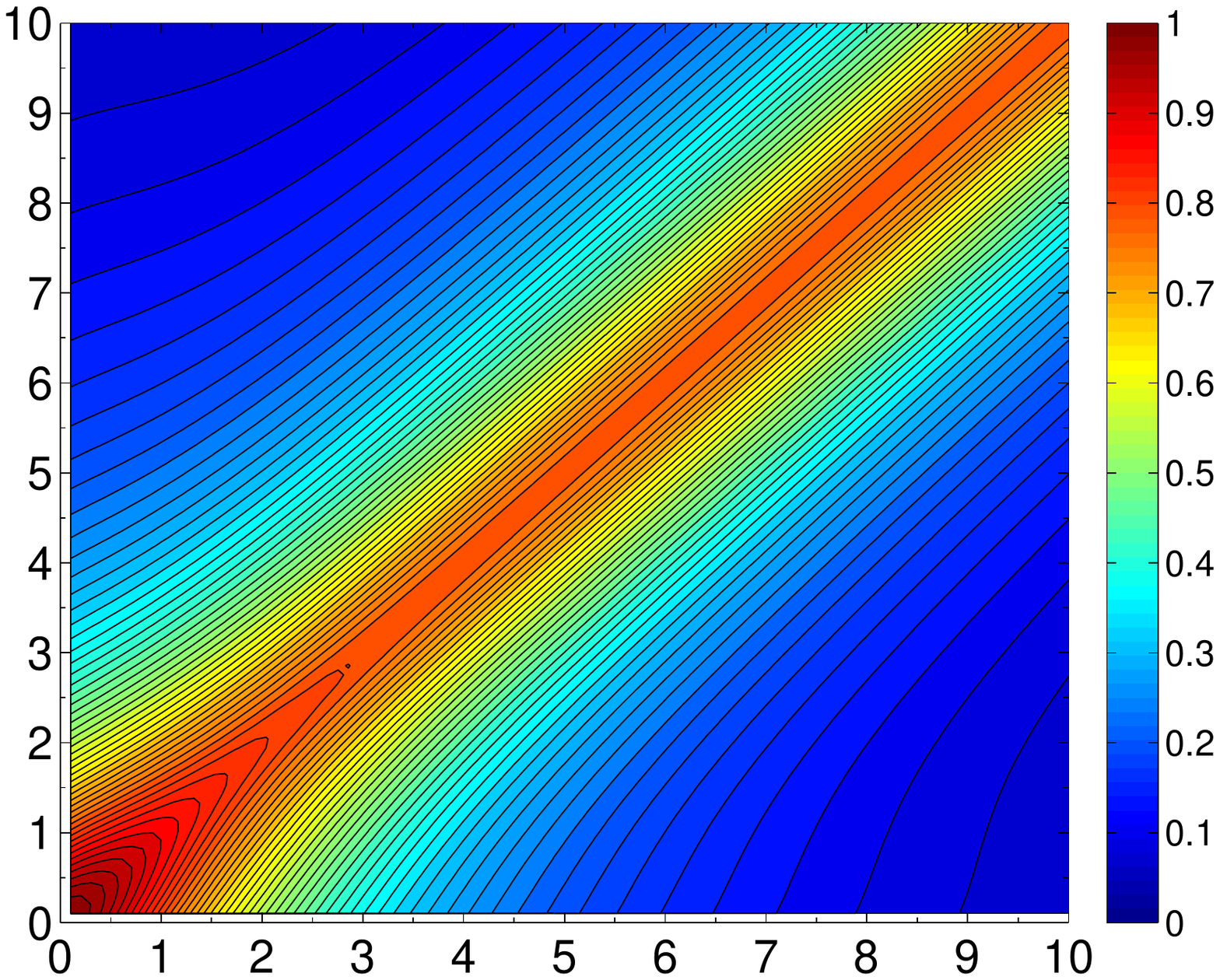}
\put(10,70){\rotatebox{90}{$(t-t^\prime)/\tau_p$}}
\put(87,3){$(t-t^{\prime\prime})/\tau_p$}\end{overpic}}
\subfloat[]
{\begin{overpic}
[trim = 20mm 55mm 0mm 50mm,scale=0.35,clip,tics=20]{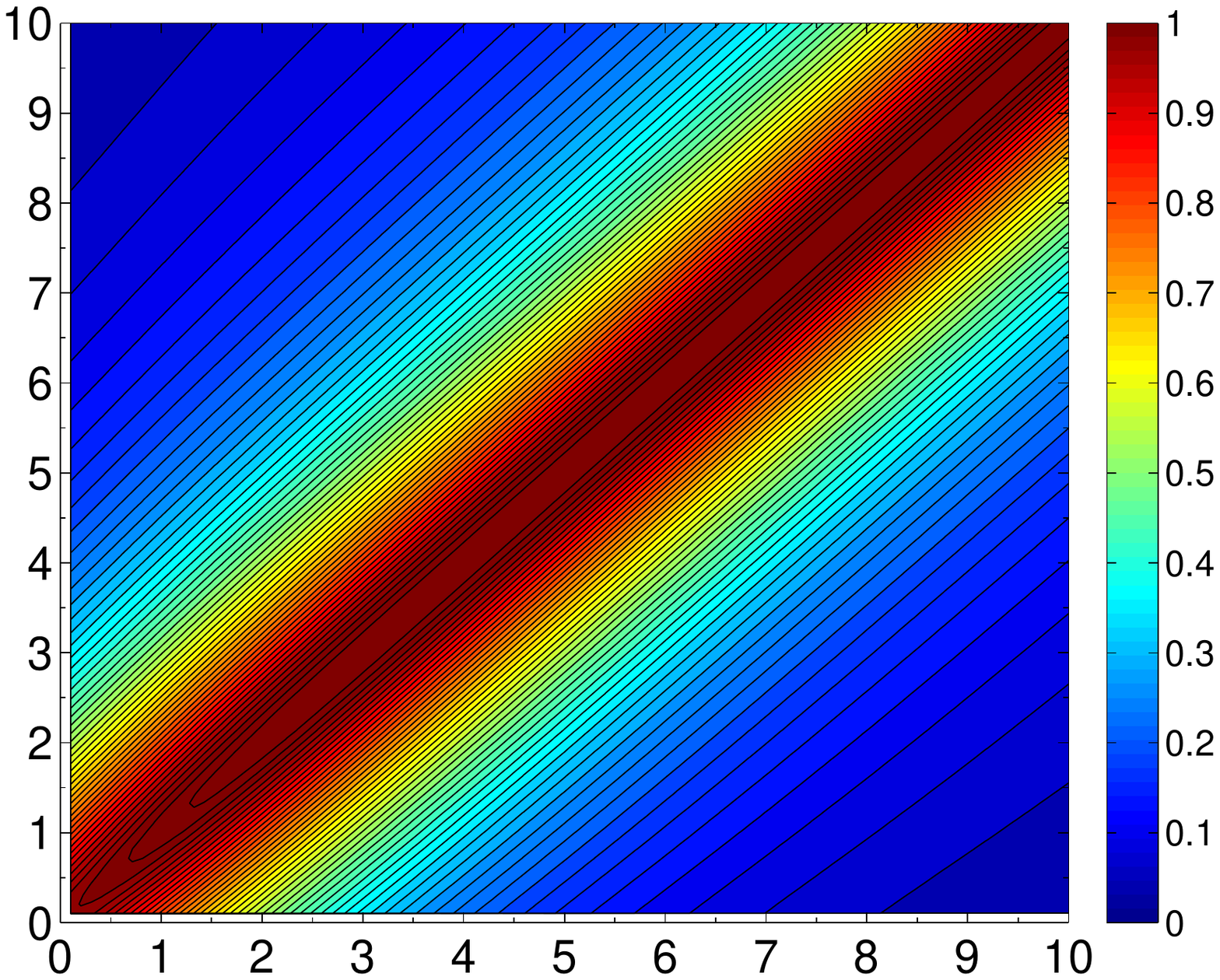}
\put(68,3){$(t-t^{\prime\prime})/\tau_p$}
\end{overpic}}
\caption{Contour plots of DNS data at ${St=1}$ for (a) ${\langle\mathcal{Z}^p(t^\prime,t^{\prime\prime})\rangle_{\mathcal{Z}^{[-]}}/\langle\mathcal{Z}^p(t,t)\rangle_{\mathcal{Z}^{[-]}}}$, (b) ${\langle\mathcal{Z}^p(t^\prime,t^{\prime\prime})\rangle_{\mathcal{Z}^{[+]}}/\langle\mathcal{Z}^p(t,t)\rangle_{\mathcal{Z}^{[+]}}}$, (c) ${\langle\mathcal{Y}^p(t^\prime,t^{\prime\prime})\rangle_{\mathcal{Z}^{[-]}}/\langle\mathcal{Y}^p(t,t)\rangle_{\mathcal{Z}^{[-]}}}$ and (d) ${\langle\mathcal{Y}^p(t^\prime,t^{\prime\prime})\rangle_{\mathcal{Z}^{[+]}}/\langle\mathcal{Y}^p(t,t)\rangle_{\mathcal{Z}^{[+]}}}$.}
\label{Autocorr} 
\end{figure}
%
\section{Conclusions}

By means of theoretical analysis and tests using DNS data we have demonstrated that the non-local clustering mechanism acting in the dissipation range of turbulence with finite $\tau_{\mathcal{S}}$ and $\tau_{\mathcal{R}}$ is consistent with the phenomenon of preferential concentration.  This occurs because the particles preferentially sample regions of high-strain and low-rotation along their path-history, and this preferential sampling affects the non-local contributions to the drift velocity generating the clustering.  The ${St= O(1)}$ regime is quite distinct from the ${St\ll1}$ regime, where preferential concentration is easily understood because the mechanism generating the clustering arises from the \emph{local} preferential sampling of the fluid velocity gradient field.
 
 For a flow with ${\tau_{\mathcal{S}}\to0}$ and ${\tau_{\mathcal{R}}\to0}$, such as in the model flow field used in \cite{gustavsson11b}, or in Navier-Stokes turbulence with particles settling under strong gravity, the centrifuge mechanism no longer operates and preferential concentration vanishes. Nevertheless, inertial particles in such a system cluster because of the non-local, path-history symmetry breaking mechanism.
 
\section*{Acknowledgements} 
 
The authors acknowledge financial support from the National Science Foundation through Grant CBET-0967349 and through the Graduate Research Fellowship awarded to PJI.  Computational simulations were performed on Yellowstone \cite{yellowstone} (ark:/85065/d7wd3xhc)
at the U.S. National Center for Atmospheric Research through its Computational and Information Systems
Laboratory (sponsored by the National Science Foundation).


\bibliographystyle{jfm}
\bibliography{refs_co12}

\begin{thebibliography}{36}
\expandafter\ifx\csname natexlab\endcsname\relax\def\natexlab#1{#1}\fi

\bibitem[Balachandar \& Eaton(2010)]{balachandar10}
{\sc Balachandar, S. \& Eaton, J.~K.} 2010 Turbulent dispersed multiphase flow.
  {\em Annu. Rev. Fluid Mech.\/} {\bf 42}, 111--133.

\bibitem[Bec(2003)]{bec03}
{\sc Bec, J.} 2003 Fractal clustering of inertial particles in random flows.
  {\em Phys. Fluids\/} {\bf 15}, L81--L84.

\bibitem[Bec {\em et~al.\/}(2007)Bec, Biferale, Cencini, Lanotte, Musacchio \&
  Toschi]{bec07}
{\sc Bec, J., Biferale, L., Cencini, M., Lanotte, A.~S., Musacchio, S. \&
  Toschi, F.} 2007 Heavy particle concentration in turbulence at dissipative
  and inertial scales. {\em Phys. Rev. Lett.\/} {\bf 98}, 084502.

\bibitem[Bec {\em et~al.\/}(2010)Bec, Biferale, Cencini, Lanotte \&
  Toschi]{bec10a}
{\sc Bec, J., Biferale, L., Cencini, M., Lanotte, A.~S. \& Toschi, F.} 2010
  Intermittency in the velocity distribution of heavy particles in turbulence.
  {\em J. Fluid Mech.\/} {\bf 646}, 527--536.

\bibitem[Bec {\em et~al.\/}(2014)Bec, Homann \& Ray]{bec14}
{\sc Bec, J., Homann, H. \& Ray, S.~S.} 2014 Gravity-driven enhancement of
  heavy particle clustering in turbulent flow. {\em Phys. Rev. Lett.\/} {\bf
  112}, 184501.

\bibitem[Bragg \& Collins(2014)]{bragg14b}
{\sc Bragg, A.D. \& Collins, L.R.} 2014 New insights from comparing statistical
  theories for inertial particles in turbulence: {I}. spatial distribution of
  particles. {\em New J. Phys.\/} {\bf 16}, 055013.

\bibitem[{Bragg} {\em et~al.\/}(2014){Bragg}, {Ireland} \& {Collins}]{bragg14a}
{\sc {Bragg}, A.~D., {Ireland}, P.~J. \& {Collins}, L.~R.} 2014 {Forward and
  backward in time dispersion of fluid and inertial particles in isotropic
  turbulence}. {\em ArXiv e-prints\/} .

\bibitem[Chong {\em et~al.\/}(1990)Chong, Perry \& Cantwell]{chong90}
{\sc Chong, M.~S., Perry, A.~E. \& Cantwell, B.~J.} 1990 A general
  classification of three-dimensional flow fields. {\em Phys. Fluids A\/} {\bf
  2}~(5), 765--777.

\bibitem[Chun {\em et~al.\/}(2005)Chun, Koch, Rani, Ahluwalia \&
  Collins]{chun05}
{\sc Chun, J., Koch, D.~L., Rani, S., Ahluwalia, A. \& Collins, L.~R.} 2005
  Clustering of aerosol particles in isotropic turbulence. {\em J. Fluid
  Mech.\/} {\bf 536}, 219--251.

\bibitem[{Computational and Information Systems Laboratory}(2012)]{yellowstone}
{\sc {Computational and Information Systems Laboratory}} 2012 Yellowstone:
  {IBM} i{D}ata{P}lex {S}ystem ({U}niversity {C}ommunity {C}omputing).
  http://n2t.net/ark:/85065/d7wd3xhc.

\bibitem[Duncan {\em et~al.\/}(2005)Duncan, Mehlig, \"{O}stlund \&
  Wilkinson]{duncan05}
{\sc Duncan, K., Mehlig, B., \"{O}stlund, S. \& Wilkinson, M.} 2005 Clustering
  by mixing flows. {\em Phys. Rev. Lett.\/} {\bf 95}, 240602.

\bibitem[Eaton \& Fessler(1994)]{eaton94}
{\sc Eaton, J.~K. \& Fessler, J.~R.} 1994 Preferential concentration of
  particles by turbulence. {\em Int. J. Multiphase Flow\/} {\bf 20}, 169--209.

\bibitem[Frisch(1995)]{frisch}
{\sc Frisch, Uriel} 1995 {\em Turbulence: {T}he Legacy of {A}. {N}.
  {K}olmogorov\/}. Cambridge University Press.

\bibitem[Girimaji \& Pope(1990)]{gir90}
{\sc Girimaji, S.~S. \& Pope, S.~B.} 1990 A diffusion model for velocity
  gradients in turbulence. {\em Phys. Fluids A\/} {\bf 2}, 242--256.

\bibitem[Gustavsson \& Mehlig(2011)]{gustavsson11b}
{\sc Gustavsson, K. \& Mehlig, B.} 2011 Ergodic and non-ergodic clustering of
  inertial particles. {\em Eur. Phys. Lett.\/} {\bf 96}, 60012.

\bibitem[Gustavsson \& Mehlig(2014)]{gustavsson14}
{\sc Gustavsson, K. \& Mehlig, B.} 2014 Clustering of particles falling in a
  turbulent flow. {\em Phys. Rev. Lett.\/} {\bf 112}, 214501.

\bibitem[Ireland {\em et~al.\/}(2014)Ireland, Bragg \& Collins]{ireland14}
{\sc Ireland, P.J., Bragg, A.D. \& Collins, L.R.} 2014 The effect of {R}eynolds
  number on inertial particle dynamics in isotropic turbulence. {P}art {I}:
  {S}imulations without gravitational effects. {\em J. Fluid Mech.\/} .

\bibitem[Ireland {\em et~al.\/}(2013)Ireland, Vaithianathan, Sukheswalla, Ray
  \& Collins]{ireland13}
{\sc Ireland, P.~J., Vaithianathan, T., Sukheswalla, P.~S., Ray, B. \& Collins,
  L.~R.} 2013 Highly parallel particle-laden flow solver for turbulence
  research. {\em Comput. Fluids\/} {\bf 76}, 170--177.

\bibitem[Ishihara {\em et~al.\/}(2009)Ishihara, Gotoh \& Kaneda]{ishihara09}
{\sc Ishihara, Takashi, Gotoh, Toshiyuki \& Kaneda, Yukio} 2009 Study of
  high-{R}eynolds-number isotropic turbulence by direct numerical simulation.
  {\em Annu. Rev. Fluid Mech.\/} {\bf 41}, 165--180.

\bibitem[Maxey(1987)]{maxey87}
{\sc Maxey, M.~R.} 1987 The gravitational settling of aerosol particles in
  homogeneous turbulence and random flow fields. {\em J. Fluid Mech.\/} {\bf
  174}, 441--465.

\bibitem[Maxey \& Corrsin(1986)]{maxey86}
{\sc Maxey, M.~R. \& Corrsin, S.} 1986 Gravitational settling of aerosol
  particles in randomly oriented cellular flow fields. {\em J. Aerosol. Sci.\/}
  {\bf 43}, 1112--1134.

\bibitem[Maxey \& Riley(1983)]{maxey83}
{\sc Maxey, M.~R. \& Riley, J.~J.} 1983 Equation of motion for a small rigid
  sphere in a nonuniform flow. {\em Phys. Fluids\/} {\bf 26}, 883--889.

\bibitem[{McQuarrie}(1976)]{mcquarrie}
{\sc {McQuarrie}, D.~A.} 1976 {\em Statistical Mechanics\/}. New York: Harper
  \& Row.

\bibitem[Pope(2000)]{pope}
{\sc Pope, S.~B.} 2000 {\em Turbulent Flows\/}. New York: Cambridge University
  Press.

\bibitem[Ray \& Collins(2013)]{ray13}
{\sc Ray, B. \& Collins, L.~R.} 2013 Investigation of sub-kolmogorov inertial
  particle pair dynamics in turbulence using novel satellite particle
  simulations. {\em J. Fluid Mech.\/} {\bf 720}, 192--211.

\bibitem[Rouson \& Eaton(2001)]{rouson01}
{\sc Rouson, D. W.~I. \& Eaton, J.~K.} 2001 On the preferential concentration
  of solid particles in turbulent channel flow. {\em J. Fluid Mech.\/} {\bf
  428}, 149--169.

\bibitem[Salazar \& Collins(2012{\natexlab{{\em a\/}}})]{salazar12b}
{\sc Salazar, J. P. L.~C. \& Collins, L.~R.} 2012{\natexlab{{\em a\/}}}
  Inertial particle acceleration statistics in turbulence: effects of
  filtering, biased sampling and flow topology. {\em Phys. Fluids\/} {\bf 24},
  083302.

\bibitem[Salazar \& Collins(2012{\natexlab{{\em b\/}}})]{salazar12a}
{\sc Salazar, J. P. L.~C. \& Collins, L.~R.} 2012{\natexlab{{\em b\/}}}
  Inertial particle relative velocity statistics in homogeneous isotropic
  turbulence. {\em J. Fluid Mech.\/} {\bf 696}, 45--66.

\bibitem[Squires \& Eaton(1991)]{squires91a}
{\sc Squires, K.~D. \& Eaton, J.~K.} 1991 Preferential concentration of
  particles by turbulence. {\em Phys. Fluids A\/} {\bf 3}, 1169--1178.

\bibitem[Sundaram \& Collins(1997)]{sundaram4}
{\sc Sundaram, S. \& Collins, L.~R.} 1997 Collision statistics in an isotropic,
  particle-laden turbulent suspension {I}. {D}irect numerical simulations. {\em
  J. Fluid Mech.\/} {\bf 335}, 75--109.

\bibitem[Wang \& Maxey(1993)]{wang93}
{\sc Wang, L.~P. \& Maxey, M.~R.} 1993 Settling velocity and concentration
  distribution of heavy particles in homogeneous isotropic turbulence. {\em J.
  Fluid Mech.\/} {\bf 256}, 27--68.

\bibitem[Wang {\em et~al.\/}(2000)Wang, Wexler \& Zhou]{wwz00}
{\sc Wang, L.-P., Wexler, A.~S. \& Zhou, Y.} 2000 Statistical mechanical
  description and modeling of turbulent collision of inertial particles. {\em
  J. Fluid Mech.\/} {\bf 415}, 117--153.

\bibitem[Wilkinson \& Mehlig(2005)]{wilkinson05}
{\sc Wilkinson, M \& Mehlig, B} 2005 Caustics in turbulent aerosols. {\em
  Europhys. Lett.\/} {\bf 71}, 186--192.

\bibitem[Zaichik \& Alipchenkov(2003)]{zaichik03}
{\sc Zaichik, L.~I. \& Alipchenkov, V.~M.} 2003 Pair dispersion and
  preferential concentration of particles in isotropic turbulence. {\em Phys.
  Fluids\/} {\bf 15}, 1776--1787.

\bibitem[Zaichik \& Alipchenkov(2007)]{zaichik07}
{\sc Zaichik, L.~I. \& Alipchenkov, V.~M.} 2007 Refinement of the probability
  density function model for preferential concentration of aerosol particles in
  isotropic turbulence. {\em Phys. Fluids\/} {\bf 19}, 113308.

\bibitem[Zaichik \& Alipchenkov(2009)]{zaichik09}
{\sc Zaichik, L.~I. \& Alipchenkov, V.~M.} 2009 Statistical models for
  predicting pair dispersion and particle clustering in isotropic turbulence
  and their applications. {\em New J. Phys.\/} {\bf 11}, 103018.

\end{thebibliography}


\begin{thebibliography}{0}
\expandafter\ifx\csname natexlab\endcsname\relax\def\natexlab#1{#1}\fi

\end{thebibliography}


\begin{thebibliography}{14}
\expandafter\ifx\csname natexlab\endcsname\relax\def\natexlab#1{#1}\fi

\bibitem[Batchelor(1971)]{Batchelor59}
{\sc Batchelor, G.~K.} 1971 Small-scale variation of convected quantities like
  temperature in turbulent fluid. part 1. general discussion and the case of
  small conductivity. {\em J.~Fluid Mech.\/} {\bf 5}, 113--133.

\bibitem[Brownell \& Su(2004)]{Brownell04}
{\sc Brownell, C.~J. \& Su, L.~K.} 2004 Planar measurements of differential
  diffusion in turbulent jets. {\em AIAA Paper 2004-2335\/}.

\bibitem[Brownell \& Su(2007)]{Brownell07}
{\sc Brownell, C.~J. \& Su, L.~K.} 2007 Scale relations and spatial spectra in
  a differentially diffusing jet. {\em AIAA Paper 2007-1314\/}.

\bibitem[Dennis(1985)]{Dennis85}
{\sc Dennis, S. C.~R.} 1985 {Compact explicit finite difference approximations
  to the Navier--Stokes equation}. In {\em Ninth Intl Conf. on Numerical
  Methods in Fluid Dynamics\/} (ed. Soubbaramayer \& J.~P. Boujot), {\em
  Lecture Notes in Physics\/}, vol. 218, pp. 23--51. Springer.

\bibitem[Hwang \& Tuck(1970)]{Hwang70}
{\sc Hwang, L.-S. \& Tuck, E.~O.} 1970 On the oscillations of harbours of
  arbitrary shape. {\em J.~Fluid Mech.\/} {\bf 42}, 447--464.

\bibitem[Koch(1983)]{Koch83}
{\sc Koch, W.} 1983 Resonant acoustic frequencies of flat plate cascades. {\em
  J.~Sound Vib.\/} {\bf 88}, 233--242.

\bibitem[Lee(1971)]{Lee71}
{\sc Lee, J.-J.} 1971 Wave-induced oscillations in harbours of arbitrary
  geometry. {\em J.~Fluid Mech.\/} {\bf 45}, 375--394.

\bibitem[Linton \& Evans(1992)]{Linton92}
{\sc Linton, C.~M. \& Evans, D.~V.} 1992 The radiation and scattering of
  surface waves by a vertical circular cylinder in a channel. {\em Phil.\
  Trans.\ R. Soc.\ Lond.\/} {\bf 338}, 325--357.

\bibitem[Martin(1980)]{Martin80}
{\sc Martin, P.~A.} 1980 On the null-field equations for the exterior problems
  of acoustics. {\em Q.~J. Mech.\ Appl.\ Maths\/} {\bf 33}, 385--396.

\bibitem[Miller(1991)]{Miller91}
{\sc Miller, P.~L.} 1991 Mixing in high schmidt number turbulent jets. PhD
  thesis, California Institute of Technology.

\bibitem[Rogallo(1981)]{Rogallo81}
{\sc Rogallo, R.~S.} 1981 Numerical experiments in homogeneous turbulence. {\em
  Tech. Rep.\/} 81835. NASA Tech.\ Mem.

\bibitem[Ursell(1950)]{Ursell50}
{\sc Ursell, F.} 1950 Surface waves on deep water in the presence of a
  submerged cylinder i. {\em Proc.\ Camb.\ Phil.\ Soc.\/} {\bf 46}, 141--152.

\bibitem[{van Wijngaarden}(1968)]{Wijngaarden68}
{\sc {van Wijngaarden}, L.} 1968 On the oscillations near and at resonance in
  open pipes. {\em J.~Engng Maths\/} {\bf 2}, 225--240.

\bibitem[Worster(1992)]{Worster92}
{\sc Worster, M.~G.} 1992 {The dynamics of mushy layers}. In {\em In
  Interactive dynamics of convection and solidification\/} (ed. S.~H. Davis,
  H.~E. Huppert, W.~Muller \& M.~G. Worster), pp. 113--138. Kluwer.

\end{thebibliography}

\end{document}